\documentclass[preprint,12pt,3p]{elsarticle}
\biboptions{sort&compress}

\usepackage{amssymb}
\usepackage{amsmath}
\usepackage{placeins}
\usepackage[table,xcdraw]{xcolor}
\usepackage{multirow}
\usepackage{booktabs}
\usepackage{caption}
\usepackage{subcaption}
\usepackage{listings}
\definecolor{codegreen}{rgb}{0,0.6,0}
\definecolor{codegray}{rgb}{0.5,0.5,0.5}
\definecolor{codepurple}{rgb}{0.58,0,0.82}
\definecolor{backcolour}{rgb}{0.95,0.95,0.92}

\lstdefinestyle{mystyle}{
    backgroundcolor=\color{backcolour},   
    commentstyle=\color{codegreen},
    keywordstyle=\color{magenta},
    numberstyle=\tiny\color{codegray},
    stringstyle=\color{codepurple},
    basicstyle=\ttfamily,
    breakatwhitespace=false,         
    breaklines=true,     
    captionpos=t,   
    keepspaces=true,     
    numbers=left,   
    numbersep=5pt, 
    showspaces=false,    
    showstringspaces=false,
    showtabs=false, 
    tabsize=2
}

\newcommand{\R}{\textsuperscript{\textregistered}}
\newcommand{\C}{\textsuperscript{\textcopyright} }

\journal{}

\begin{document}

\begin{frontmatter}




\title{Optimal shape design of printing nozzles for extrusion-based additive manufacturing}

\author[inst2,inst3,inst4]{Tom\'as Schuller}
\author[inst7]{Maziyar Jalaal}
\author[inst6]{Paola Fanzio}
\author[inst4,inst5]{Francisco J. Galindo-Rosales \corref{cor1} }

\cortext[cor1]{Corresponding author}

\affiliation[inst2]{organization={Institute of Science and Innovation in Mechanical and Industrial Engineering (INEGI)},
            addressline={Rua Dr. Roberto Frias, 400}, 
            city={Porto},
            postcode={4200-465},
            country={Portugal}}

\affiliation[inst3]{organization={Transport Phenomena Research Center (CEFT), Mechanical Engineering Department},
            addressline={Faculty of Engineering of the University of Porto, Rua Dr. Roberto Frias s/n}, 
            city={Porto},
            postcode={4200-465}, 
            country={Portugal}}

\affiliation[inst4]{organization={ALiCE—Associate Laboratory in Chemical Engineering},
            addressline={Faculty of Engineering of the University of Porto, Rua Dr. Roberto Frias s/n}, 
            city={Porto},
            postcode={4200-465}, 
            country={Portugal}}

\affiliation[inst7]{organization={Van der Waals-Zeeman Institute, Institute of Physics, University of Amsterdam},
            addressline={Science Park 904}, 
            city={Amsterdam},
            postcode={1098XH}, 
            country={The Netherlands}}

\affiliation[inst6]{organization={Department of Precision and Microsystems Engineering (PME), Faculty of Mechanical, Maritime and Materials Engineering (3mE)},
            addressline={TU Delft (Delft University of Technology), Mekelweg 2}, 
            city={Delft},
            postcode={2628 CD}, 
            country={The Netherlands}}

\affiliation[inst5]{organization={Transport Phenomena Research Center (CEFT), Chemical Engineering Department},
            addressline={Faculty of Engineering of the University of Porto, Rua Dr. Roberto Frias s/n}, 
            city={Porto},
            postcode={4200-465}, 
            state={},
            country={Portugal}}            
            

\begin{abstract}
The optimal design seeks the best possible solution(s) for a mechanical structure, device, or system, satisfying a series of requirements and leading to the best performance. In this work, optimized nozzle shapes have been designed for a wide range of polymer melts to be used in extrusion-based additive manufacturing, which aims to minimize pressure drop and allow greater flow control at large extrusion velocities.
This is achieved with a twofold approach,  combining a global optimization algorithm with computational fluid dynamics for optimizing a contraction geometry for viscoelastic fluids and validating these geometries experimentally.
In the optimization process, variable coordinates for the nozzle's contraction section are defined, the objective function is selected, and the optimization algorithm is guided within manufacturing constraints. Comparisons of flow-type and streamline plots reveal that the nozzle shape significantly influences flow patterns. Depending on the rheological properties, the optimized solution either promotes shear or extensional flow, enhancing the material flow rate. Finally, experimental validation of the nozzle performance assessed the actual printing flow, the extrusion force and the overall print control. It is shown that optimizing the nozzle can significantly reduce backflow-related pressure drop, positively impacting total pressure drop (up to 41 \%) and reducing backflow effects. This work has real-world implications for the additive manufacturing industry, offering opportunities for increased printing speeds, enhanced productivity, and improved printing quality and reliability. Our research contributes to advancing extrusion-based printing processes technology, addressing industry demands and enhancing the field of additive manufacturing.
\end{abstract}

\begin{graphicalabstract}
\includegraphics[width=\textwidth]{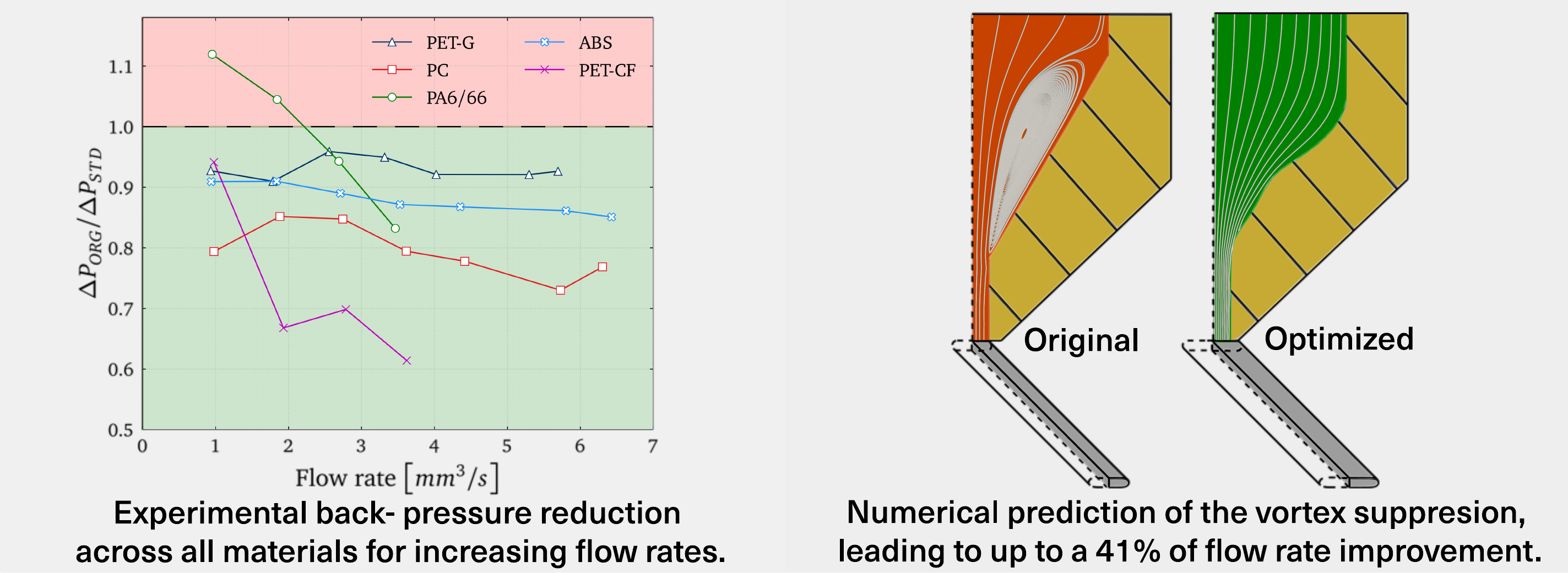}
\end{graphicalabstract}




\end{frontmatter}


\FloatBarrier

\section{Introduction}

Extrusion-based additive manufacturing, such as Fused Filament Fabrication (FFF), has revolutionized the field of additive manufacturing, enabling the creation of complex 3D structures with ease and affordability. The printing nozzle is at the heart of the FFF process, being a critical component responsible for precise material deposition~\cite{OSSWALD201851,GAO2021101658,Yadav2022}. When it is not performing as expected, this behavior can lead to print quality issues~\cite{turner,turner2}.

The extrusion process has a fundamental role in defining the print quality, and it is one of the most significant factors in limiting the maximum printing speed~\cite{larsson_nozzle,Bikas2016}. 
Mechanically, the molten polymer undergoes an axisymmetric contraction flowing through the nozzle. This flow configuration induces a complex deformation rate distribution in the extruded material, being purely elongational in the centerline of the geometry, shear-dominated near the walls and a combination of both at any location in between~\cite{POOLE2023105106}. This results in an excess pressure drop that depends not only on the geometric design of the contraction but also on the rheological properties of the extruded material~\cite{PEREZCAMACHO2015260,sietsepaper,paper2}.
Rothstein and McKinley conducted comprehensive studies of contraction flows using highly viscous Boger fluids with low Reynolds numbers, emphasizing elasticity's impact on pressure drops \cite{rothstein,rothstein1}. They found that the pressure drop increased with the formation of an upstream vortex due to elastic instabilities. Microfluidic experiments have revealed similar pressure drops, even with planar flow geometries, and showcased upstream circulatory flows at high flow rates and Deborah numbers \cite{james1982planar,nigen2002viscoelastic,rodd2005inertio,campo2011flow}. 
We have recently reported that upstream vortices indeed occur in FFF, where the upstream vortex tip marks the culmination of elastic stresses due to extensional flow~\cite{schuller_add_ma}. Moreover, the rheological properties of the fluids also impact the shape of the vortex and the dependence of the pressure drop with the flow rate through the nozzle~\cite{paper2}. This pressure drop affects the equilibrium height in the backflow region, according to Equation~\ref{eq:experimental}~\cite{schuller_add_ma,sietsepaper}:

\begin{equation}
\label{eq:experimental}
 H^{\star} = \frac{\Delta \textrm{P}_{tot}-\Delta \textrm{P}_{n}}{(N_{1-BF} + \tau_{w-BF}) \cdot \frac{4}{D_f}},
\end{equation}

\noindent where $ \Delta \textrm{P}_{tot}$ is the total pressure, $\Delta \textrm{P}_{n}$ is the nozzle pressure, $D_f$ is the filament diameter, $N_{1-BF}$ is the shear-induced normal stress difference, and $\tau_{w-BF}$ is the shear stress at the wall of the filament in the backflow region.

James and Roos recently demonstrated~\cite{james2021} that a viscoelastic fluid can flow through a converging channel and not generate upstream vortices, leading to a pressure drop across the contraction identical to that of a Newtonian fluid with the same viscosity. This result suggests that elasticity would not affect the pressure drop along the nozzle but would still affect the die swell at the exit. This is possible through a modification in the shape of the nozzle, which allows the distribution of normal stresses and pressure in the radial direction so that streamlines migrate towards the centerline, reducing the shear rate and shear stress at the wall. Consequently, the energy spent for shearing in the channel would be reduced.
Their channel geometry, consisting of an axisymmetric hyperbolic channel with a conical section at the inlet (total angle of 60\textdegree), was designed to avoid an upstream vortex and to maximize extensional effects, specifically, to achieve Deborah numbers up to 3 along with a minimum strain of 3. However, their dimensions (total length of 122~mm and 3~mm exit die diameter) were impractical for FFF applications. These findings, along with our previous results \cite{sietsepaper,paper2}, inspired the idea of designing nozzle shapes for specific polymer rheologies to suppress upstream vortices and minimize pressure drop. A nozzle shape with such characteristics would increase the extrusion velocity and even reduce the risk of backflow.

Optimization is a branch of Applied Mathematics that is applied across various industrial domains, including structural engineering, aerospace, automotive, chemical, electronics, and manufacturing \cite{rao2019engineering,siddall1982optimal,mdo_book,opt_cca}. Optimization involves algorithms determining the maxima and minima of objective functions linked to the best design to meet predefined objectives and geometric or system constraints. In the realm of design optimization, there are various approaches and techniques that can be employed to improve the efficiency, performance, and effectiveness of a design. Some common approaches are trial and error \cite{Galindo-Rosales2023}, sizing, and shape optimization \cite{ma12071086}. These methods are part of mathematical and heuristic optimization approaches.
Topology Optimization (TO) takes a different route, dynamically altering the design's topology throughout the optimization process. What sets TO apart from the previous approaches is its independence from the initial configuration. This allows it to arrive at optimal designs based on defined objectives and constraints \cite{C3RA47230B}.

Borvall and Petersson \cite{borrvall_petersson} are considered the first researchers to use topology optimization in fluid mechanics. Twenty years ago, the authors formulated an optimization problem to minimize the total potential power loss, which, in the absence of body forces, is reduced to the minimization of the dissipated energy in the fluid. Since then, the use of TO has steadily grown and, in recent years, it has also been a matter of research to create microfluidic geometries for different purposes, including the use of viscoelastic fluids, with remarkable results~\cite{JensenAPL2012,Groisman2004,SOUSA2010652,pimenta4,Zografosbiomic2016,C3RA47230B,zografos2019,Haward2012,liu_zografos}. The objectives of these optimized shapes are varied, but they all stem from the aim of controlling flow rate, the flow characteristics and pressure drop in small channel geometries.

This study delves into numerically optimizing 3D printing nozzle designs to reduce nozzle pressure drop in FFF to enhance material flow rate and mitigate pressure-related issues while abiding to the spatial constraints in a commercial FFF 3D printer and the manufacturing constraints. 
Optimized geometries for different materials in FFF will be fabricated to be tested experimentally. A thorough validation and analysis of the experimental results will allow us to assess whether numerically optimized nozzle geometries lead to real-world performance improvements in extrusion velocity and reduce backflow issues and how this optimized shape may influence print quality and reliability.

\FloatBarrier
\section{Methods}
\subsection{Polymers}


In order to have a wide representation of the most common materials used in FFF, the selected polymers for this study are PET-G, PC, PA6/66, ABS and PET-CF (carbon fiber-reinforced PET variant). This selection was made to have filaments representative of the main physical properties, that is, amorphous and crystalline materials, standard and reinforced filaments, as well as engineering grade and more general purpose plastics.
The rheological information, as well as the modeling of these polymers and subsequent testing, has already been detailed in previous works \cite{schuller_add_ma,paper2}.

\subsection{Numerical optimization}

\subsubsection{Flow simulation and shape optimization loop}

DAKOTA\C \cite{adams2020dakota} can be used with OpenFOAM\R\cite{foam_jasak} to perform high-fidelity simulation and optimization of complex flows, including those with turbulence, multiphase flow, and other complex physics. In this study, an engineering design loop using DAKOTA\C as an optimizer with a loosely coupled approach was applied. By using the command line interface and shell scripting, different applications, such as \textit{Gmsh}\C \cite{Geuzaine2009} and OpenFOAM\R~were integrated. The procedure for the numerical simulations in the optimization loop, the definition of the boundary conditions and the constitutive equations have already been described in previous works~\cite{schuller_add_ma,sietsepaper,paper2} and are present in an abridged version on \ref{app:const_eq}.

In our study, the objective function is the minimization of total pressure drop in the nozzle, as it is a global measure of the performance of the contraction and allows a better understanding of the force being made on the solid filament by the molten polymer.
To parameterize the contraction section, we resorted to the use of splines to model the contraction. 
They are particularly useful in parameterization because they can accurately describe complex relationships, allowing for efficient and continuous variation of parameters while maintaining smooth transitions between values. 
For this, 11 variables were defined, being them the \textit{r}-radius of the contraction at each increment of the \textit{z}-length. These variables were bounded with upper and lower limits, in which they could vary from individual to individual.

\begin{figure}[htbp]
	\begin{center}	
		\includegraphics[width=0.8\textwidth]{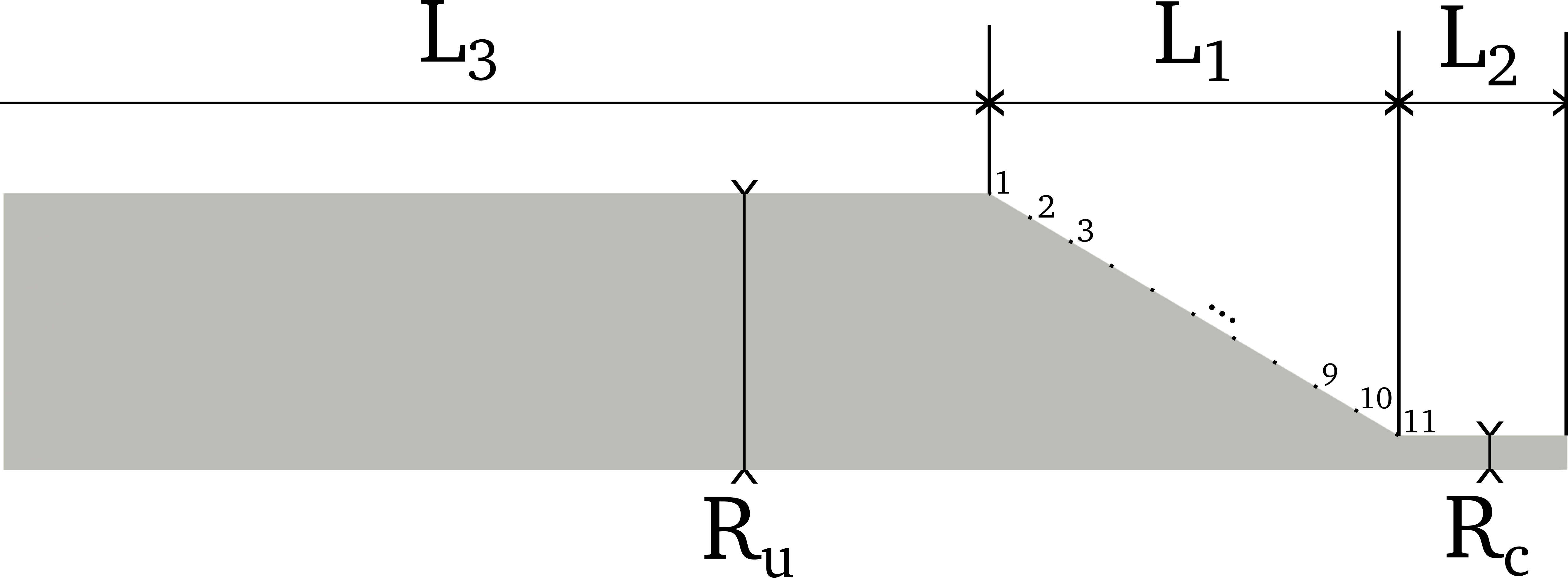}
        \caption{Standard nozzle geometry. The numbered dots demonstrate the optimized points on the spline.}
		\label{fig:bb_dims}
	\end{center}
\end{figure}

\noindent The design space is based around the five characteristic dimensions, two radii  ($R_u$, $R_c$) and three lengths ($L_1$, $L_2$ and $L_3$), that are featured in the standard nozzle geometry (see Figure \ref{fig:bb_dims} and Table \ref{tab:base_opt_dims_table}). The initial positions of the 11 points of the spline are also present in Fig. \ref{fig:bb_dims}.

\begin{table}[htbp]
\centering
\caption{Base nozzle dimensions.}
\resizebox{0.4\textwidth}{!}{%
\begin{tabular}{cc}
\toprule \textbf{Dimensions}      &  \textbf{Standard nozzle }    \\  \midrule
$L1$       &    2.425~mm     \\ 
$L2$       &    1~mm         \\ 
$L3$       &    30~mm        \\ 
$R_u$      &    1.6~mm          \\ 
$R_c$      &    0.2~mm        
 \\\bottomrule
\end{tabular}}
\label{tab:base_opt_dims_table}
\end{table}

\begin{figure}[htbp]
\centering
  \begin{subfigure}[b]{\linewidth}
    \centering
    \includegraphics[width=0.6\textwidth]{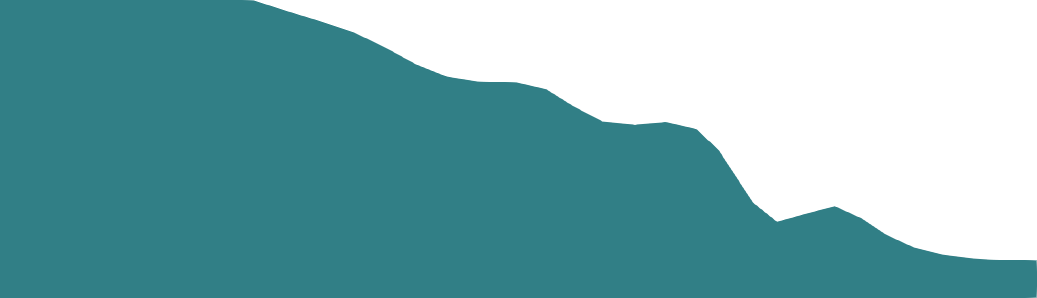}
	\caption{Gradient-descent optimized geometry.}
	\label{fig:gradient_first}
\end{subfigure}
\begin{subfigure}[b]{\linewidth}
    \centering
    \includegraphics[width=0.6\textwidth]{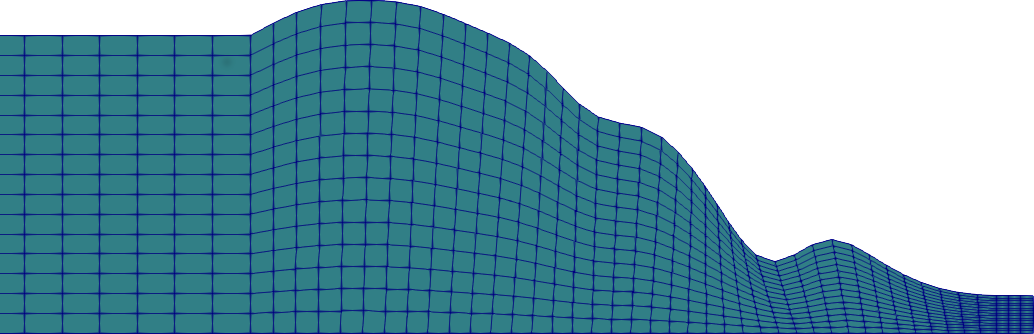}
		\caption{Final SOGA-optimized mesh.}
		\label{fig:soga_final}
\end{subfigure}
 \caption{Optimized geometries.}
 \label{fig:opt_geos}
\end{figure}

To evaluate the robustness of the loop and reach benchmark results, first, an unconstrained optimization process was used to reduce the pressure drop when working with Polycarbonate (PC) with an 
extrusion velocity of $\overline{V_{ext}}=$ 110~mm/s, which corresponds to a volumetric flow rate larger than 10 mm$^3$/s.
A gradient-descent optimization was performed to reach a local minimum generated by the algorithm, producing a geometry that reduced the inlet pressure by 60.6~\%. This geometry is shown in Figure \ref{fig:opt_geos} a).

Using this geometry, a grid convergence study (GCS) was performed (results shown in \ref{app:a}) and the grid size was chosen, allowing reliable results and saving computational time. Then, a more in-depth optimization was performed using a Single Objective Genetic Algorithm (SOGA), a powerful global optimization routine in the DAKOTA\C optimization library. The optimization running with the variables presented in Table \ref{tab:soga_table_all} in \ref{app:b} resulted in an organic optimal nozzle shape, shown in Figure \ref{fig:opt_geos} b), that further reduces the inlet pressure by 66.8~\%.

\subsubsection{Geometrical constraints for nozzle fabrication}

\begin{figure}[htbp]
	\begin{center}		
		\includegraphics[width=\textwidth]{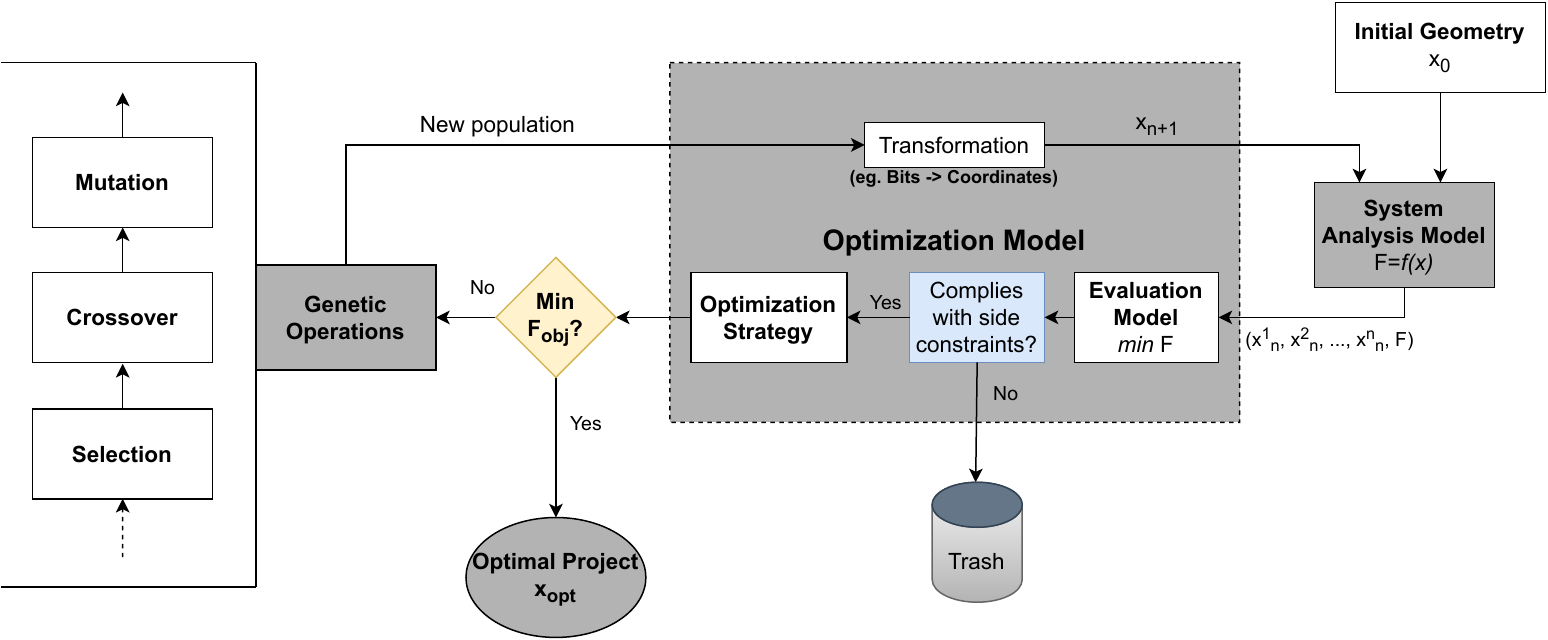}
		\caption{Constrained SOGA optimization loop, adapted from \cite{rao2019engineering}.}
		\label{fig:opt_loop_adapt}
	\end{center}
\end{figure}

The presence of protrusions in the organic optimized geometry may compromise its fabrication by means of conventional milling processes, so the algorithm was modified, and geometrical constraints were incorporated to ensure the nozzle was manufacture-ready (Figure \ref{fig:opt_loop_adapt}). These constraints ensured that the nozzle geometry would have no ``bulges'' and decrease monotonically, as this is the only type of geometry a conventional manufacturing process can produce. In the algorithm, it was specified that every downstream radius variable ($n+1$) was forcibly lower than the one before it ($n$).

This modified algorithm was run for all 5 polymers, producing 5 final shapes for manufacturing.

\FloatBarrier
\subsection{Optimized nozzle manufacturing and experimental setup}

\begin{figure}[htbp]
\centering
\begin{subfigure}[t]{.39\textwidth}
    \centering
    \includegraphics[width=\textwidth]{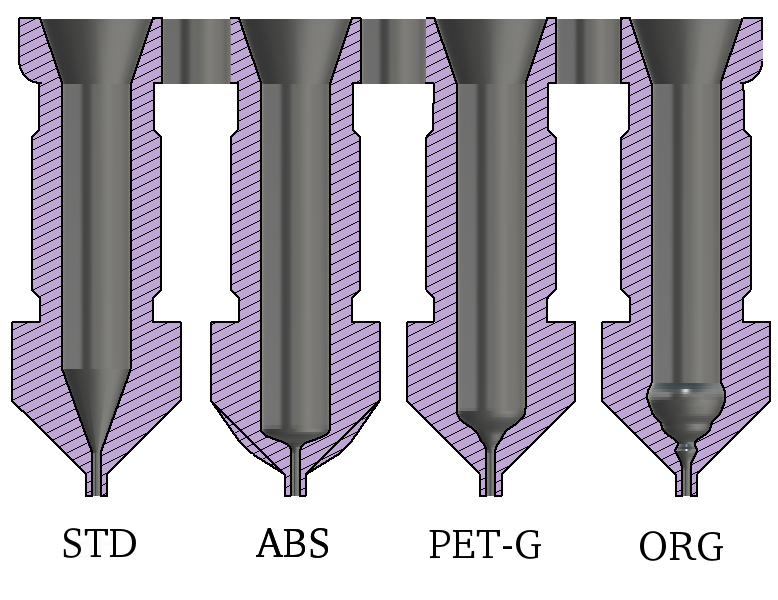}
    \caption{Nozzle sections.}
\end{subfigure}
\hfill
\begin{subfigure}[t]{.58\textwidth}
    \centering
    \includegraphics[width=\textwidth]{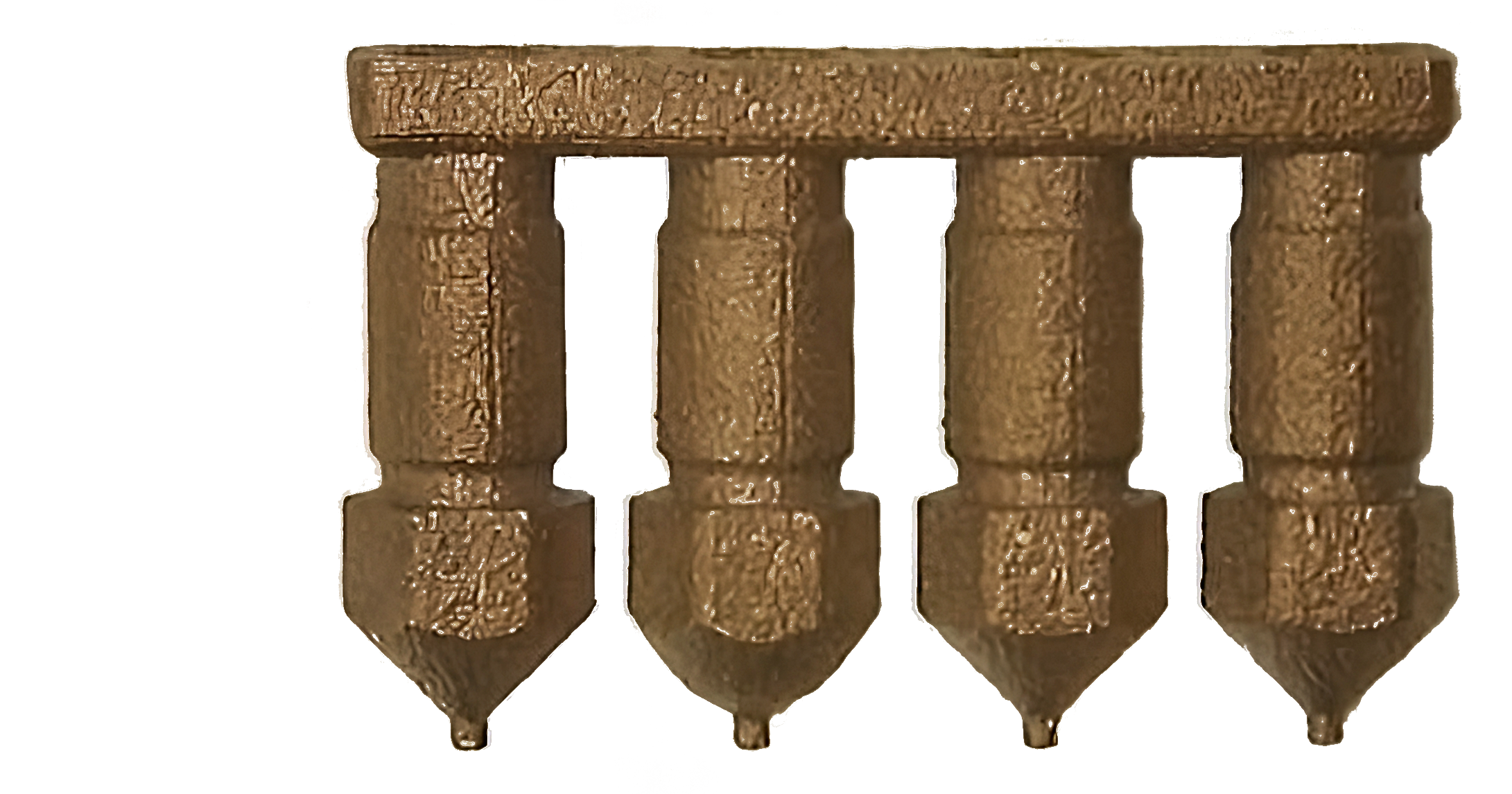}
    \caption{Metal printed nozzles.}
\end{subfigure}%
\caption[short]{Default and optimized nozzles.}
\label{fig:brassnozzles}
\end{figure}

From the 5 optimized nozzles, using a cross-validation method in which every modeled polymer was run in each geometry (results in \ref{app:e}), 2 final geometries were selected, along with the standard shape and the unconstrained organic shape (Figure~\ref{fig:brassnozzles} a). These were manufactured in brass by metal binder jetting, which enables the production of complex parts with intricate internal geometries \cite{Li2020}.
The standard nozzles were produced to allow a comparison of the results exclusively based on the shape of the nozzles. Figure~\ref{fig:brassnozzles} b) shows the printed set of nozzles.

The nozzles for the experimental campaign were post-processed in-house by cutting the thread, reaming the nozzle exit, and soldering a thermocouple for temperature measurement.  
The final post-processed nozzle and die exit dimension evaluation are featured in \ref{app:c} (Figures \ref{fig:final_nozzle} and \ref{fig:die_dims}).

\begin{figure}[htbp]
	\begin{center}		
		\includegraphics[width=0.4\textwidth]{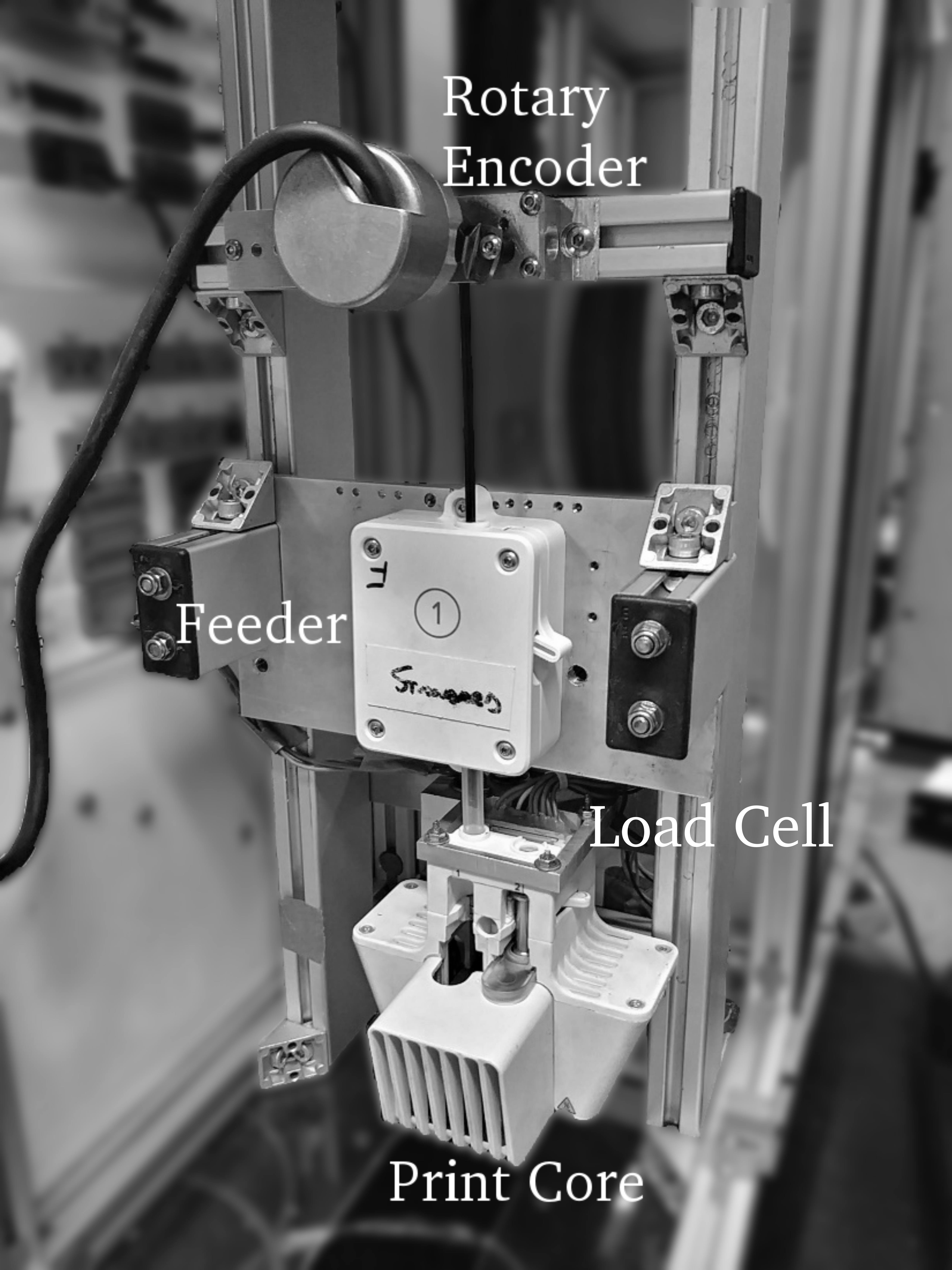}
		\caption{Free extrusion test setup.}
		\label{fig:monostrudel}
	\end{center}
\end{figure}

The extrusion experiments occurred in a static, open-air environment without printhead movement, as in previous works~\cite{schuller_add_ma,sietsepaper}. A stepper motor controlled the adjustable feeder, and to account for slipping, a rotary encoder measured the filament flow rate. The filament was guided into the liquefier and pushed into the nozzle, where it melted. The force applied by the feeder was measured using a load cell (Figure \ref{fig:monostrudel}).

Extrusion measurements were conducted by adjusting the input filament volumetric flow rate within a range of 1 mm$^3$/s to 8 mm$^3$/s. A continuous extrusion was recorded for 10 minutes, and after each test, the data was saved as a single hierarchical data format 5 (hdf5) file, with a ten-second interval separating each test. Three independent runs were performed for each experimental setting, and the average force for each test with constant temperature and feed rate was measured. We excluded the data collected in the initial 150 seconds to allow for extrusion stabilization and to reach a steady state.

\begin{figure}
  \begin{minipage}[c]{.4\linewidth}
    \centering
    \includegraphics[width=0.9\linewidth]{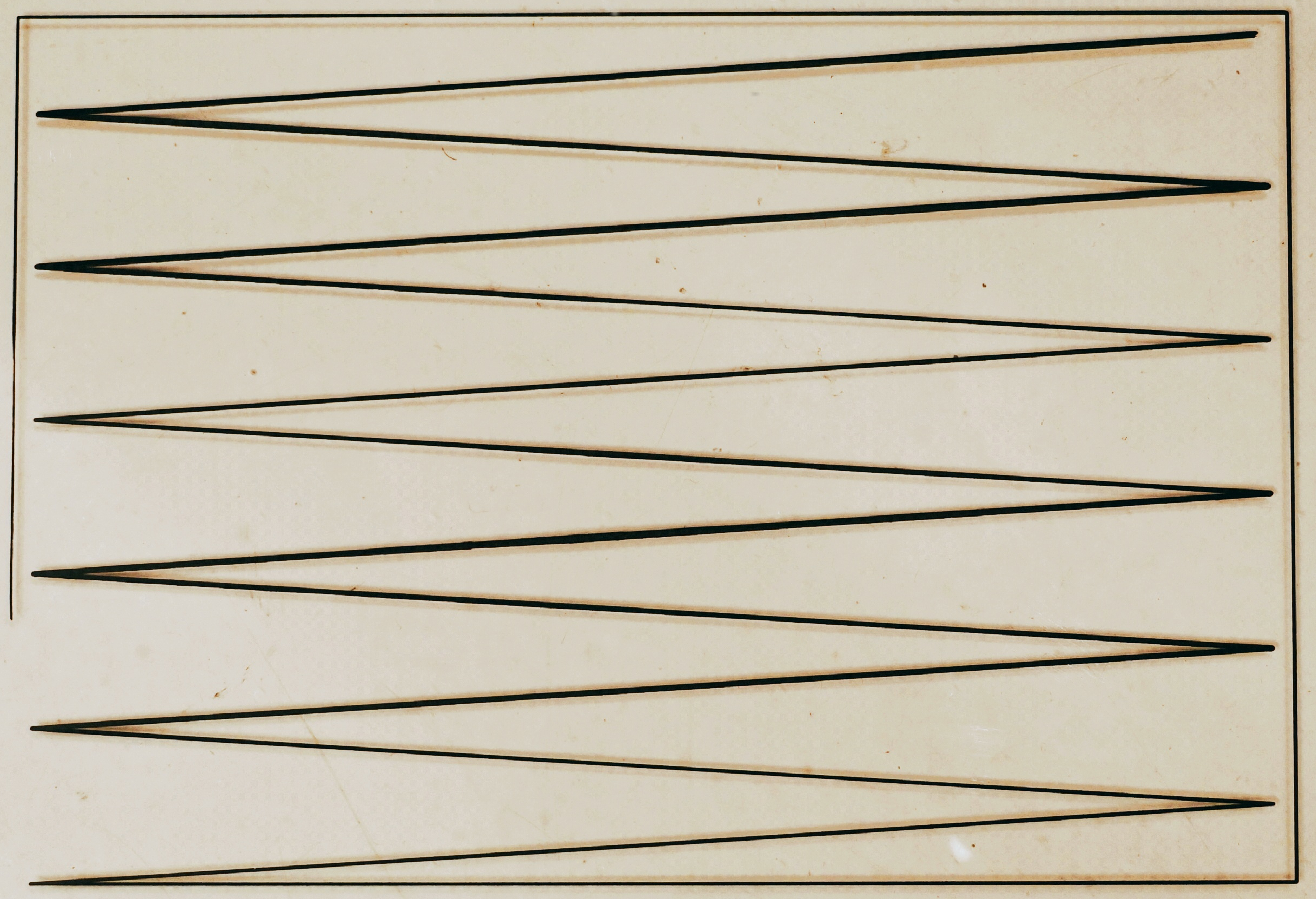}
    \captionof{figure}{Final 1-layer deposition print (PET-CF, optimized nozzle).}
    \label{fig:zigzags}
  \end{minipage}\hfill
  \begin{minipage}[c]{.6\linewidth}
    \centering
        \captionof{table}{Line deposition test parameters.}
        \resizebox{0.95\textwidth}{!}{%
\begin{tabular}{lccccc}
\toprule    \textbf{Material}          &\textbf{ PET-G  }& \textbf{PC}     & \textbf{PA6/66} & \textbf{ABS}    & \textbf{PET-CF} \\ \midrule
Flow rate     & \multicolumn{5}{c}{1, 3, 5, 8 mm$^3$/s}     \\
Print speed   & 20 mm/s & 20 mm/s & 20 mm/s & 20 mm/s & 20 mm/s \\
Layer height  & 0.1 mm  & 0.1 mm  & 0.1 mm  & 0.1 mm  & 0.1 mm  \\
Fan speed     & 50\%   & 50\%   & 50\%   & 50\%   & 1\%    \\
T$_{extrusion}$ & 240\textdegree~C  & 260\textdegree~C  & 243\textdegree~C  & 240\textdegree~C  & 265\textdegree~C  \\
T$_{bed}$       & 100\textdegree~C  & 100\textdegree~C  & 60\textdegree~C   & 85\textdegree~C   & 100\textdegree~C \\\bottomrule
\end{tabular}%
\label{tab:1line1layer}
}
  \end{minipage}
\end{figure}

Moreover, 1-layer deposition tests were performed on a stock UltiMaker\R~S5, where a single layer of material is extruded, forming single lines on the printer buildplate, using a custom script to generate the \textit{toolpath}, a zigzag pattern with flow rate changes at each line midpoint (see Figure \ref{fig:zigzags}). This test allows for assessing the nozzle performance in printing conditions while producing easily measurable results, such as the width and height of the line. The test parameters are present in Table \ref{tab:1line1layer}.

\subsection{Line Profilometry with Optical Coherence Tomography}


To evaluate the section of the printed filament in the 1-layer deposition test, Optical Coherence Tomography (OCT) was employed.
OCT utilizes low coherence interferometry to create a depth scan of a sample \cite{Huang1991} and provides a fast and simple way to obtain the surface profile with a few micron resolution \cite{vanderKolk2023}

\begin{figure}[htbp]
\centering
\begin{subfigure}{.48\textwidth}
    \centering
    \includegraphics[width=\textwidth]{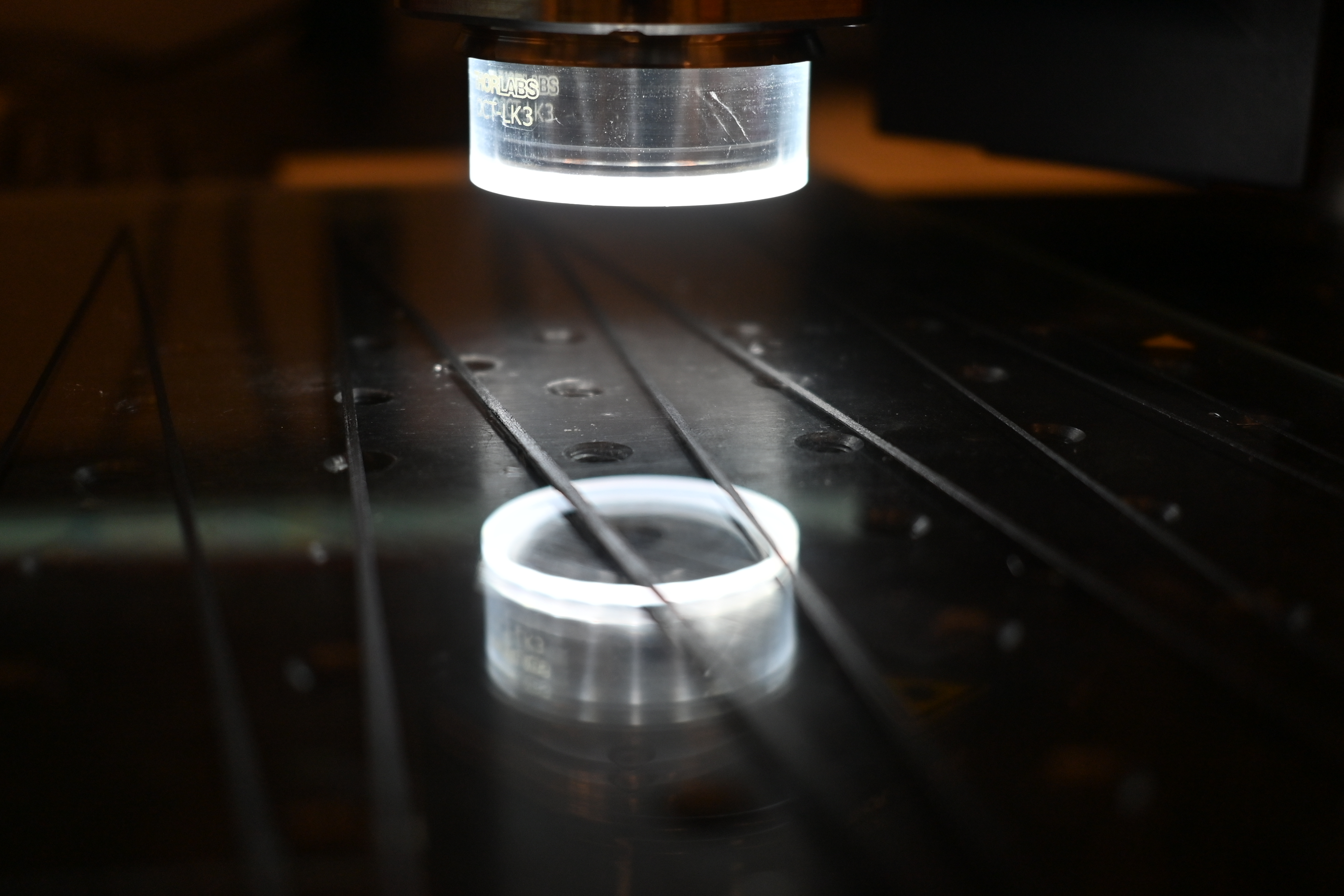}
    \caption{Scanner closeup during operation.}
\end{subfigure}%
\hfill
\begin{subfigure}{.48\textwidth}
    \centering
    \includegraphics[width=\textwidth]{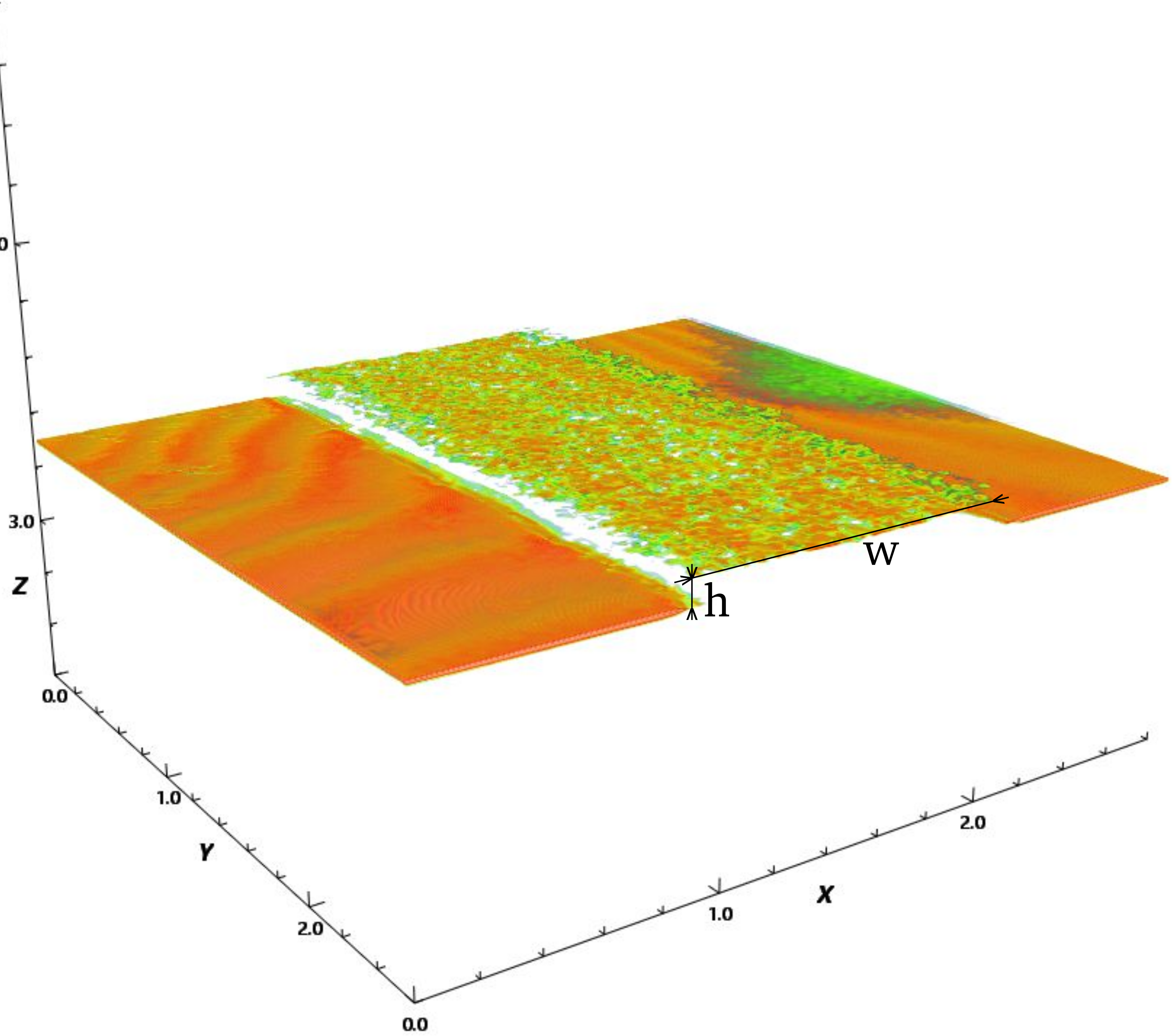}
    \caption{Isometric view of scan heightmap results.}
\end{subfigure}
\caption[short]{OCT polymer line measurement details.}
\label{fig:isoscans}
\end{figure}

In this study, the setup consisted of a Thorlabs Telesto OCT System
, with a center wavelength of 1300 nm, imaging depth (in air) of 3.5 mm, axial resolution (in air) of 5.5 $\mu$m, A-Scan line rate of 5.5, 28, 48, \& 76 kHz and sensitivity of 96 dB (at 76 kHz) to 111 dB (at 5.5 kHz). The width $w$ and height $h$ of each polymer line were measured (Figure \ref{fig:isoscans}) three times for statistical significance, which along with the prescribed print-head speed, the actual print flow rate could be calculated.

\section{Results and discussion}


\subsection{Numerical results}
\begin{figure}[ht!] 
    \centering
\includegraphics[height=.5\textwidth]{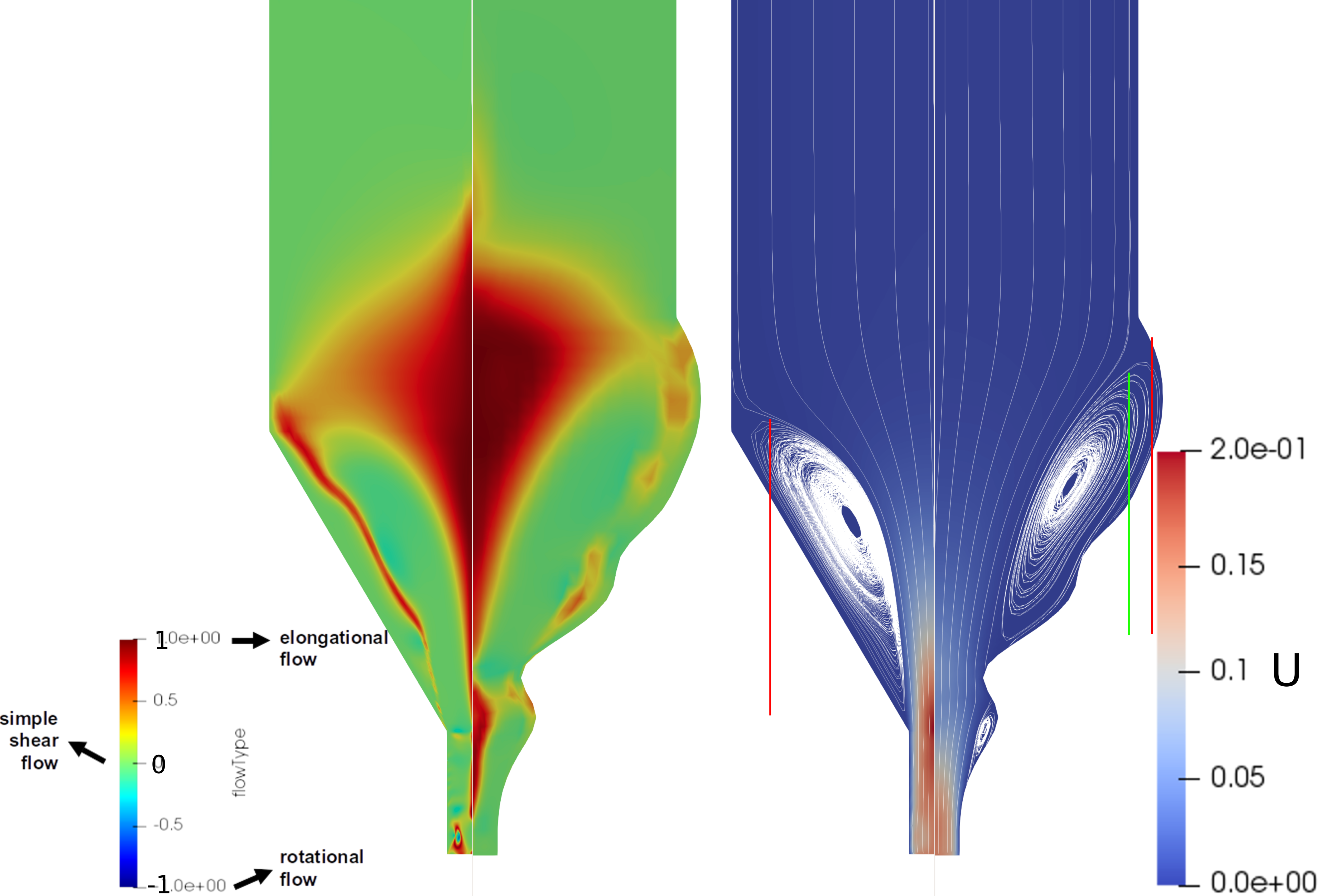}
    \caption{Flow-type, path-lines, velocity field (units in m/s) comparison between the original and the organic optimized geometries using the modeled PC.} 
 \label{fig:comparison}
\end{figure}

Numerical simulations reveal the complex flow field during the material extrusion. At the wall, due to the no-slip condition, the flow is dominated by shear flow, while the contraction leads to an elongation dominated flow close to the centerline.
Any fluid particle located in between these two regions experiences a combination of both flow types. In fact, the flow-type parameter, $\xi=\frac{\|\boldsymbol{D}\| - \|\boldsymbol{\Omega}\|}{\|\boldsymbol{D}\| + \|\boldsymbol{\Omega}\|}$ \cite{ORTEGACASANOVA2019102,schuller_add_ma}, helps in visualizing which kind of flow is dominating in each location (see Figure \ref{fig:comparison}).
According to the analysis of James~\cite{james2016}, the ratio between normal stresses is proportional to the polymeric viscosity $\eta_p$ and inversely proportional to the zero-shear first normal stress coefficient, $\Psi_{10}$, and extension rate $\dot\epsilon$ for fixed dimensions in the length of the nozzle ($L$) and the diameter of the die ($D$), i.e. $\frac{N_{1-S}}{N_{1-E}}\propto \frac{\eta_p}{\Psi_{10} \dot\epsilon}$, where $N_{1-S}$ is the shear-dominated 1\textsuperscript{st} normal stress difference, and $N_{1-E}$ is the extensional-dominated 1\textsuperscript{st} normal stress difference. Thus, for a given material, to promote $N_{1-S}$ against $N_{1-E}$, the extension rate ($\dot\epsilon\propto \frac{1}{h(z)z}$) must be decreased, which can be achieved by enlarging $h(z)$. Correcting the value of $\frac{N_{1-S}}{N_{1-E}}$ may end up in a reduction of the elastic instabilities and the extra pressure drop, ideally in a Newtonian-like behavior with the suppression of the upstream vortices~\cite{james2021}. That is the reason explaining why for the PC the organic shape presented these bulges in the tapered region (Figure~\ref{fig:comparison}). Despite not suppressing the upstream vortices, due to the limited space for the contraction ($L/D\downarrow$), the flow-type pattern was modified so that the shear flow was promoted.

\begin{figure}[h!]
\centering
\begin{subfigure}{.28\textwidth}
    \centering
    \includegraphics[width=\textwidth]{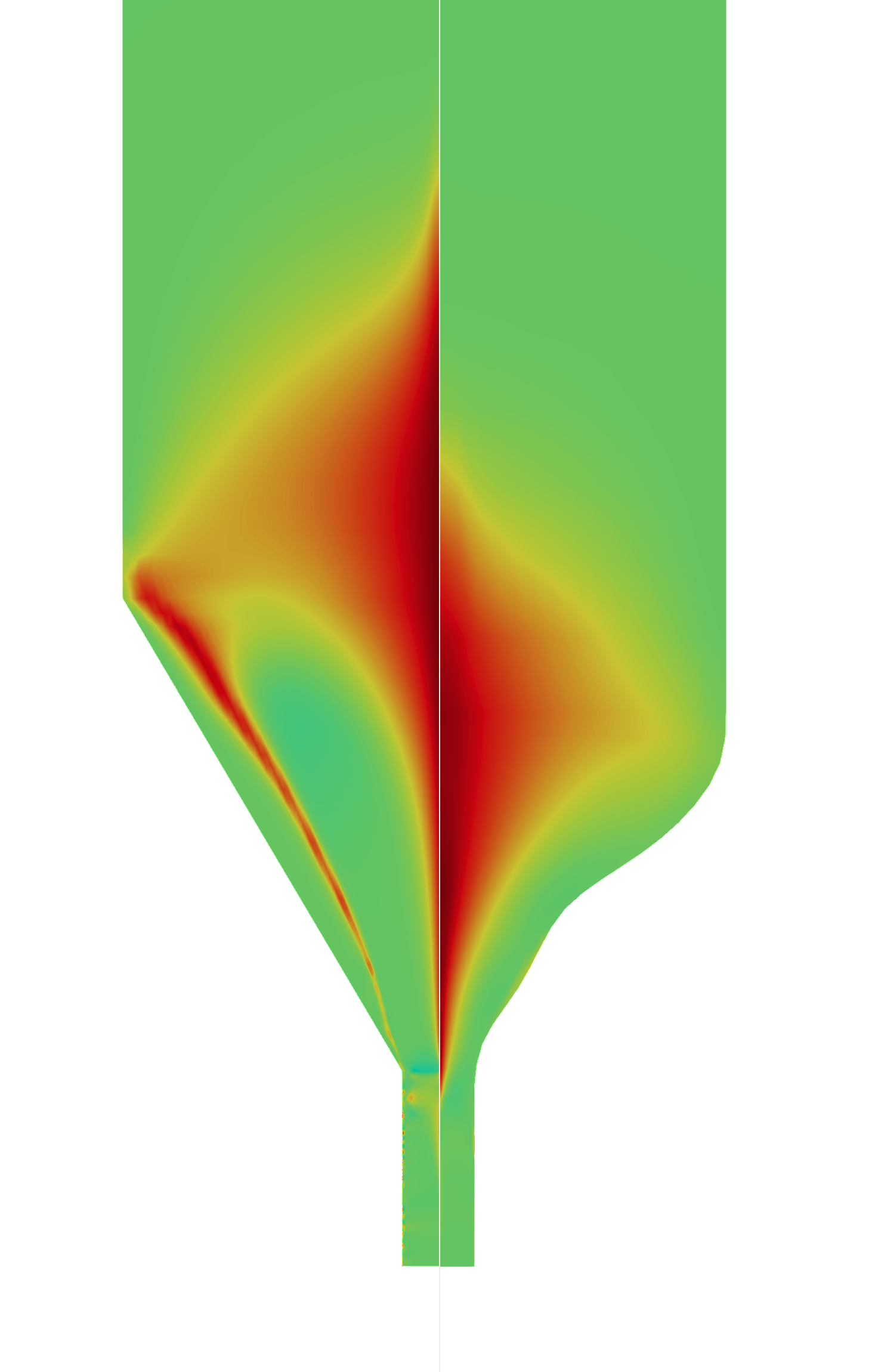}
    \caption{PET-G.}
\end{subfigure}%
\begin{subfigure}{.28\textwidth}
    \centering
    \includegraphics[width=\textwidth]{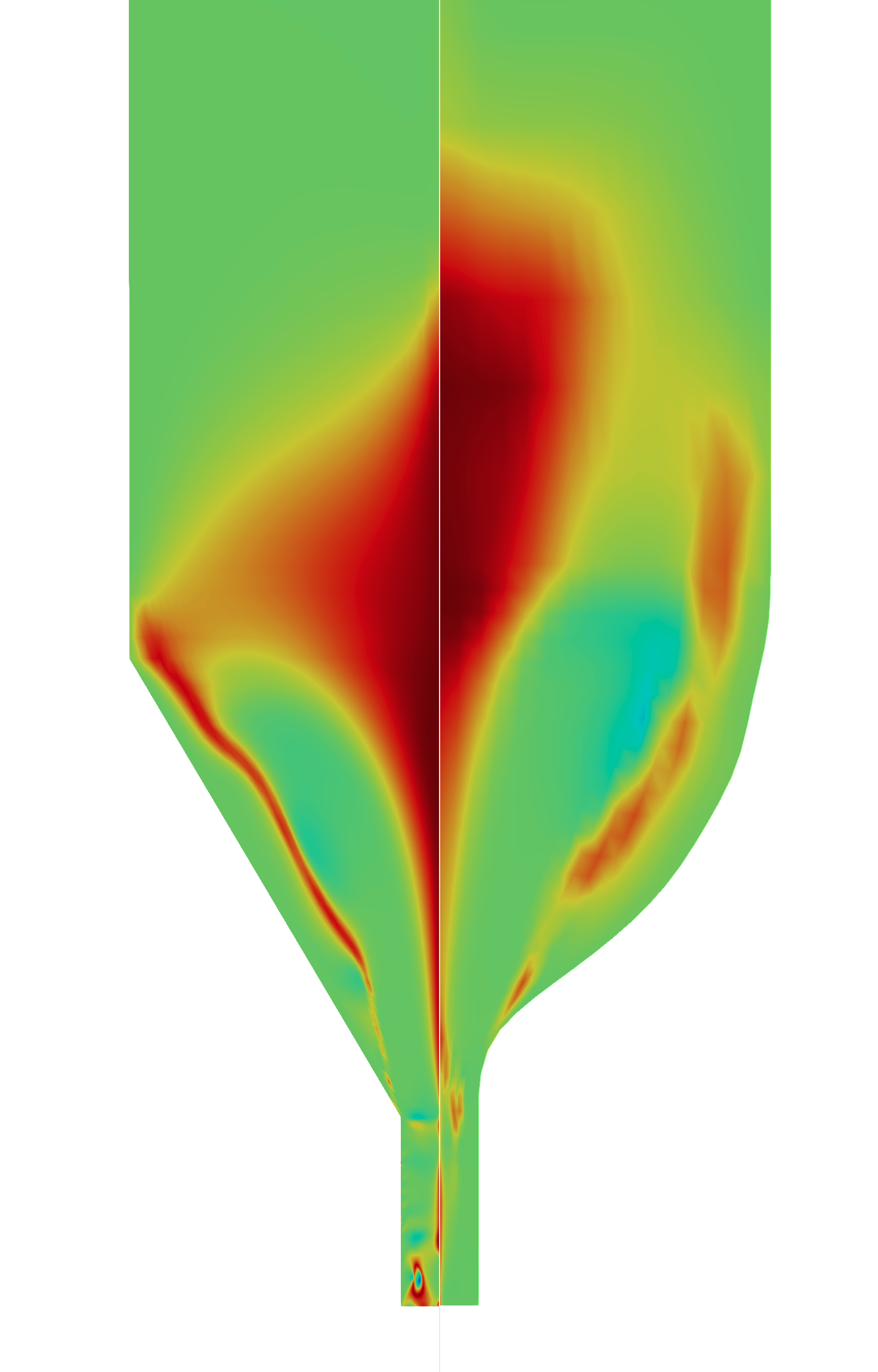}
    \caption{PC.}
\end{subfigure}
\begin{subfigure}{.28\textwidth}
    \centering
    \includegraphics[width=\textwidth]{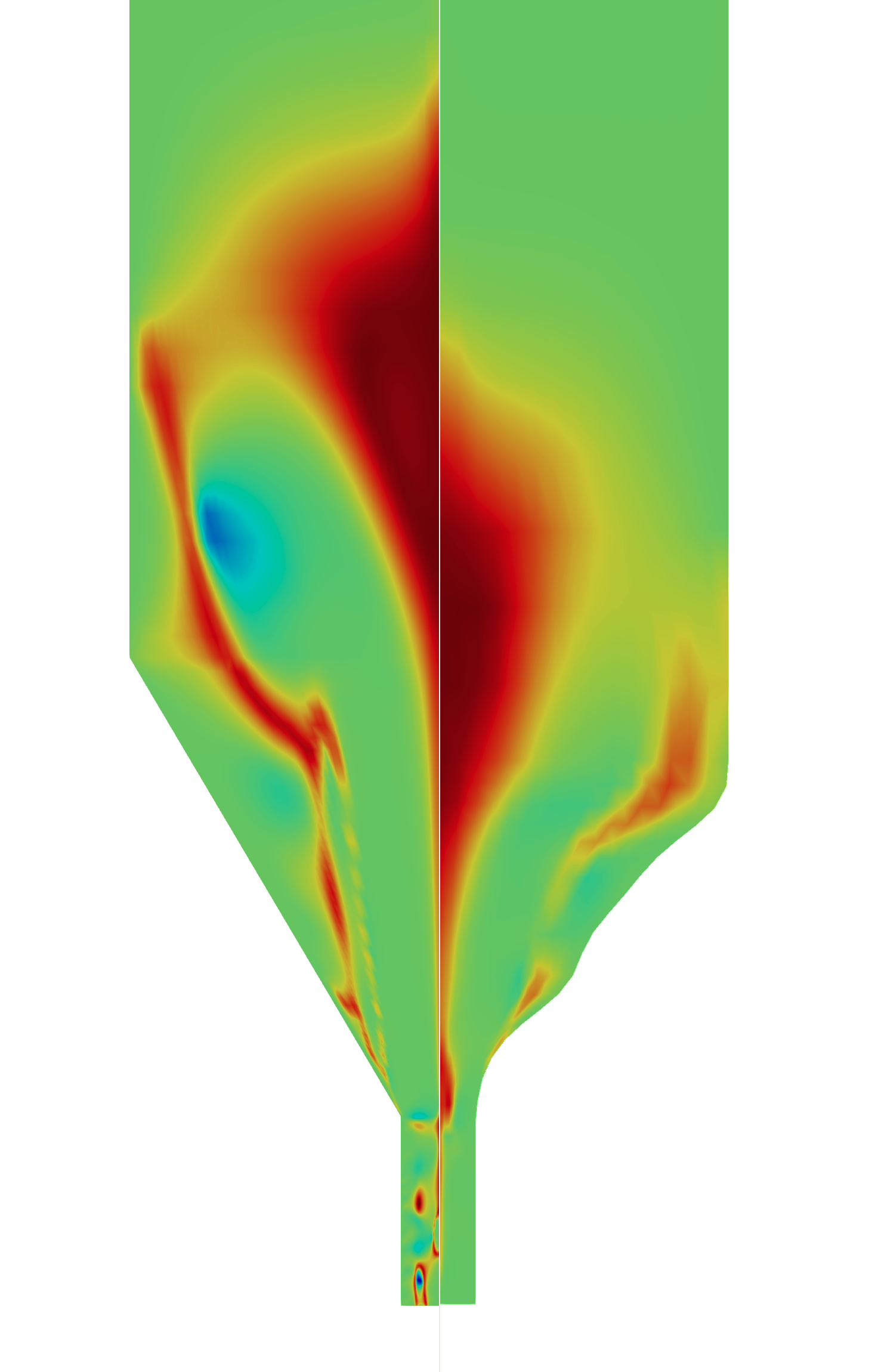}
    \caption{PA6/66.}
\end{subfigure}
\begin{subfigure}{0.315\textwidth}
    \centering
    \includegraphics[width=\textwidth]{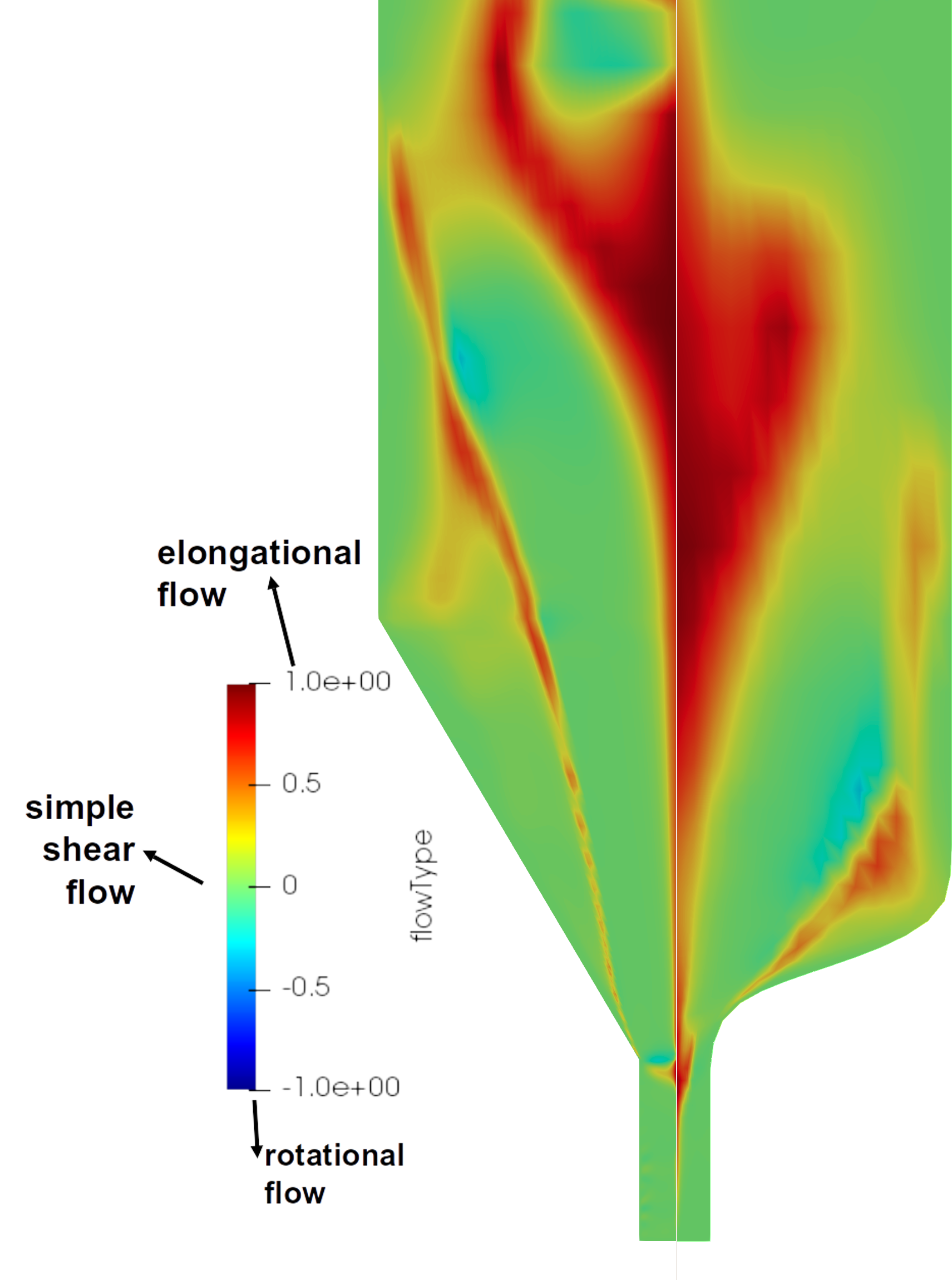}
    \caption{ABS.}
\end{subfigure}
\begin{subfigure}{.27\textwidth}
    \centering
    \includegraphics[width=\textwidth]{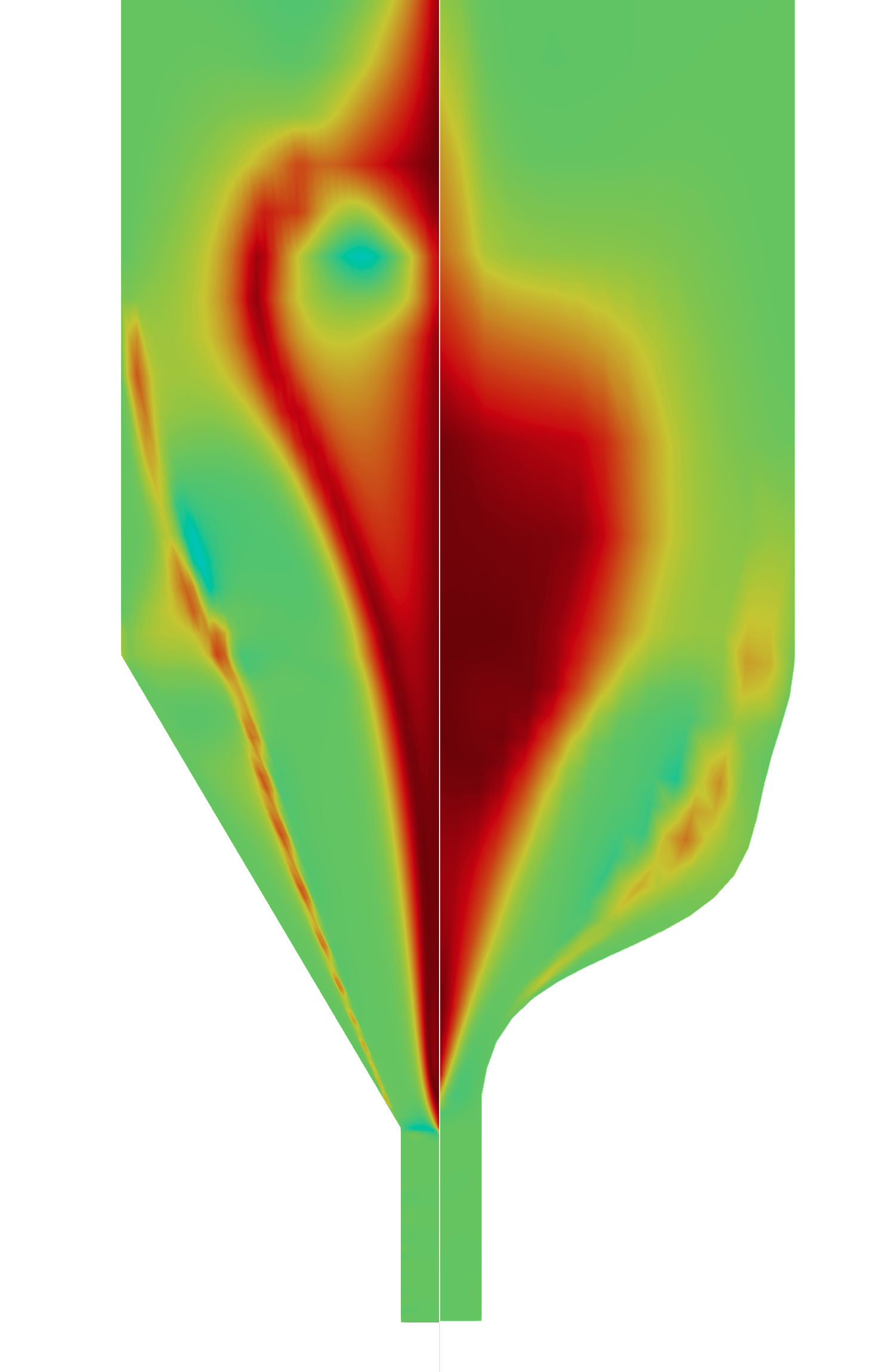}
    \caption{PET-CF.}
\end{subfigure}
\begin{subfigure}{.4\textwidth}
    \centering
    \includegraphics[width=\textwidth]{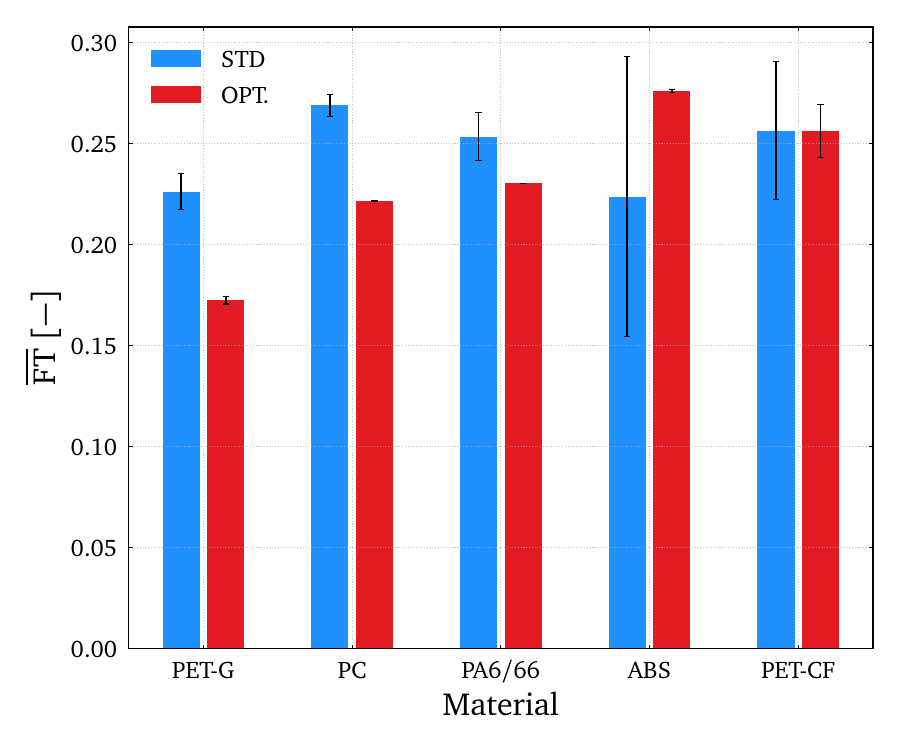}
    \caption{Average flow-type.}
\end{subfigure}
\caption{(a-e) Comparison between standard and optimized geometries in terms of flow-type pattern per material at 110 mm/s. (f) Bar plot graph comparing the average flow-type ($\overline{FT}=\frac{\int_V FTdV}{V}~\left[-\right]$) in the fluid domain for each nozzle shape and geometry, where the error bars are representative of the elastic instabilities~\cite{paper2}.}
\label{fig:flowtypecomparison}
\end{figure}

 Figures \ref{fig:flowtypecomparison} a-e) show a comparison between the flow-type contour plot of the flow of the different polymers through the original nozzle geometry and their optimized nozzle shape. Contrary to the organic optimal shape, the restrictions limited the possibility of enlarging the radius of the tapered region and changes in the extension rate were achieved by shortening or enlarging the length of the tapered region. Thus, changing the nozzle shape resulted in a different configuration of the flow-type within the fluid domain while maintaining the simple shear flow at the wall and the elongational flow at the centerline. The objective function was the minimization of the pressure drop along the nozzle, which depends on the contribution of three terms: the viscous dissipation of energy, where the larger the viscosity of the material, the larger the pressure drop; the pressure-drop reduction associated with the shear-induced normal stress ($N_{1-S}$), which help in facilitating the flow; and the extra-pressure drop associated to the extension-induced normal stresses ($N_{1-E}$)~\cite{Walters2009TheCR, Tseng2023}. Depending on the rheological properties of the extruded material, the optimal shape solution may promote either the extensional flow or the shear flow. Based on the results of our previous work \cite{paper2}, we can cluster the materials based on the relaxation time and the value of the alpha parameter: Group 1 for low $\alpha$ and low relaxation time (PET-G), Group 2 for high $\alpha$ and low relaxation time (PA6/66 and PC), and Group 3 for high relaxation time (ABS and PET-CF). Figure~\ref{fig:flowtypecomparison}.(f) shows that polymers from Groups 1 and Group 2 were provided with optimal nozzles that promote the shear flow, whereas the optimal nozzles for the polymers from Group 3, due to their large polymer viscosity, increased the extensional flow to result in a lower pressure drop. Moreover, for all cases, it can also be observed that the optimized nozzle mitigated the elastic instabilities occurring with the standard nozzle. The flow velocity and recirculation size comparisons can be found in \ref{app:e} (Figures~\ref{fig:streamlines} and \ref{fig:taus}, and Tables~\ref{tab:drop_all_mat_geos} and \ref{tab:recirc_all_mat_geos}), for all the materials.

\begin{figure}[b!]
\centering
\includegraphics[width=\linewidth]{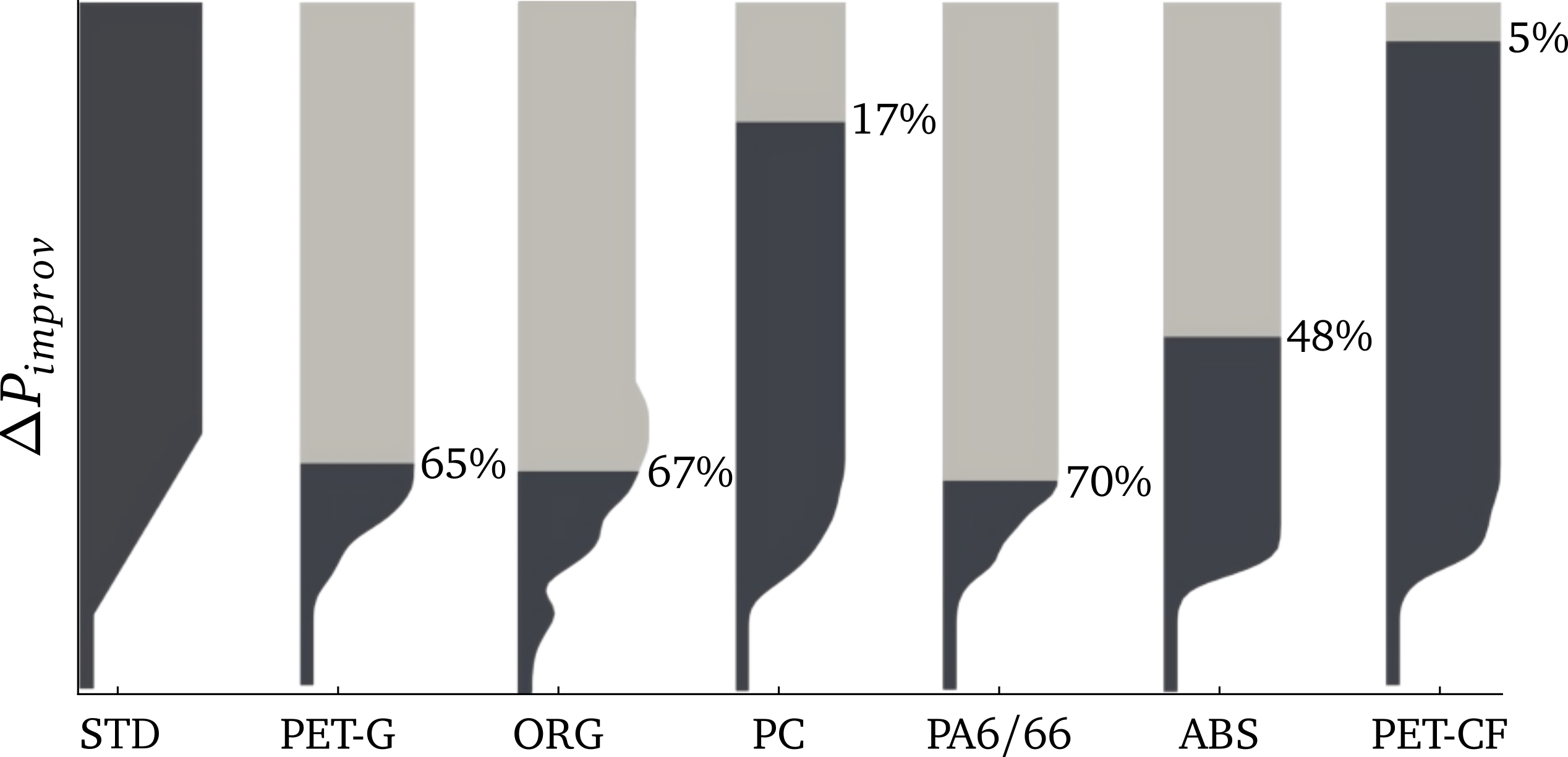}
\caption{Pressure drop reduction per geometry, being the organic one optimized also for PC.}
\label{fig:geometryimprov}
\end{figure}

The pressure drop improvements compared to the standard nozzle shape are shown in Figure~\ref{fig:geometryimprov}. As previously hypothesised \cite{paper2}, the most sensitive materials were those exhibiting larger variation on pressure drop with vortex size ($\frac{d\Delta P}{dX_r}$), i.e.  PA6/66 and PET-G, are the ones that have a greater improvement due to a geometry change, 70\% and 65\% decrease in pressure drop, respectively. In the case of PC, it is striking how imposing the design restrictions severely reduced the pressure drop improvement. Finally, it is also eye-catching that PET-CF was so insensitive to the optimization process, probably also due to the restrictions imposed on the design.

To validate these findings experimentally, four geometries have been selected and manufactured based on the numerical results presented above, i.e. one optimized nozzle for each group of polymers and the original shape. Thus,
PET-G, Organic, and ABS-optimized geometries have been chosen for representing materials from Groups 1, 2 and 3, respectively, as they have a better average improvement in pressure drop (Table~\ref{tab:drop_all_mat_geos}).

\subsection{Extrusion process}

\begin{figure}[htbp]
\centering
\begin{subfigure}[t]{.45\textwidth}
    \centering
    \includegraphics[width=\textwidth]{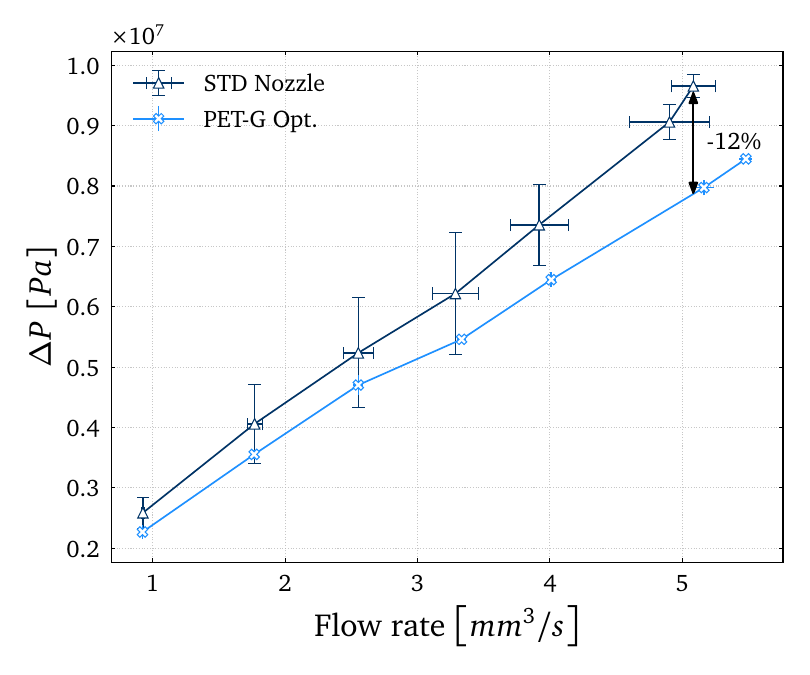}
    \caption{PET-G.}
\end{subfigure}%
\begin{subfigure}[t]{.45\textwidth}
    \centering
    \includegraphics[width=\textwidth]{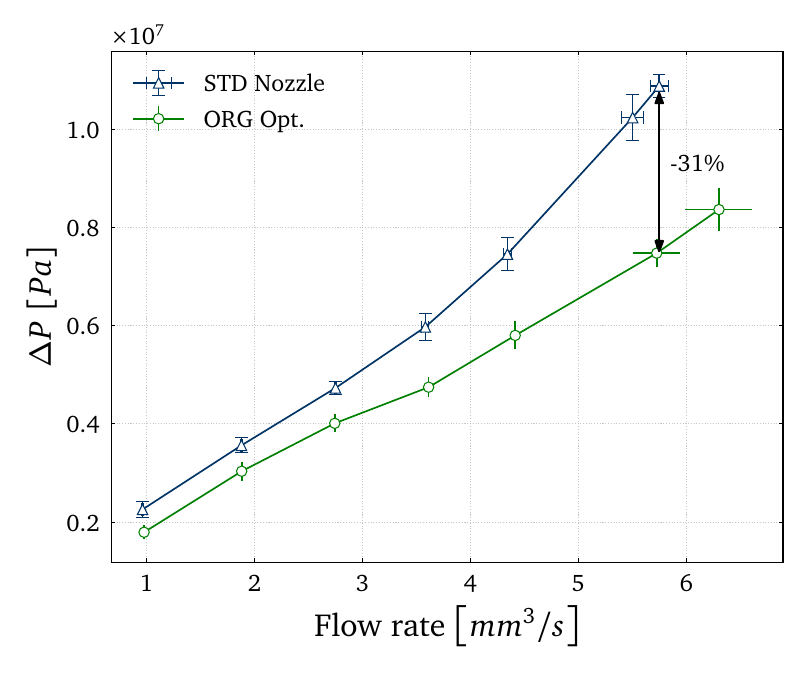}
    \caption{PC.}
\end{subfigure}
\begin{subfigure}[t]{.45\textwidth}
    \centering
    \includegraphics[width=\textwidth]{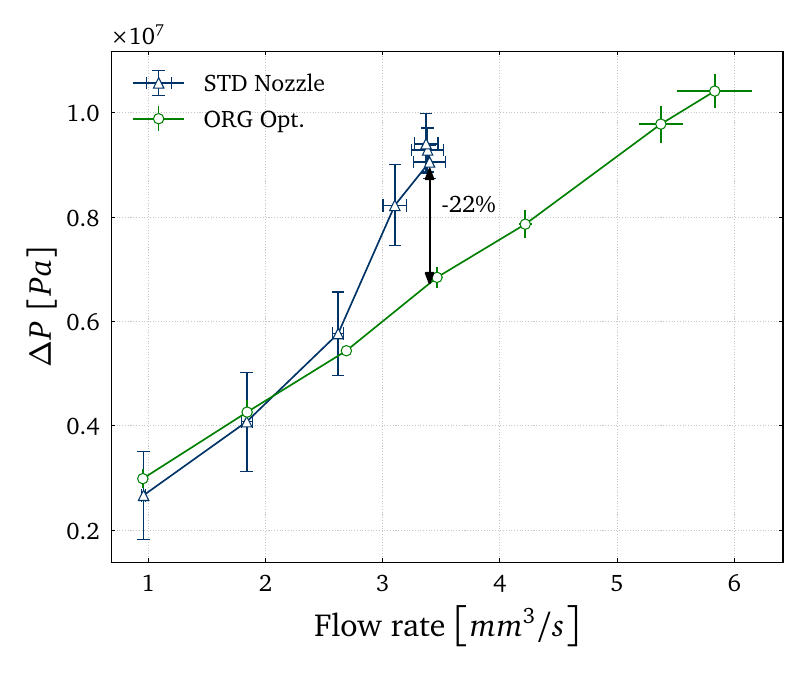}
    \caption{PA6/66.} 
\end{subfigure}
\begin{subfigure}[t]{.45\textwidth}
    \centering
    \includegraphics[width=\textwidth]{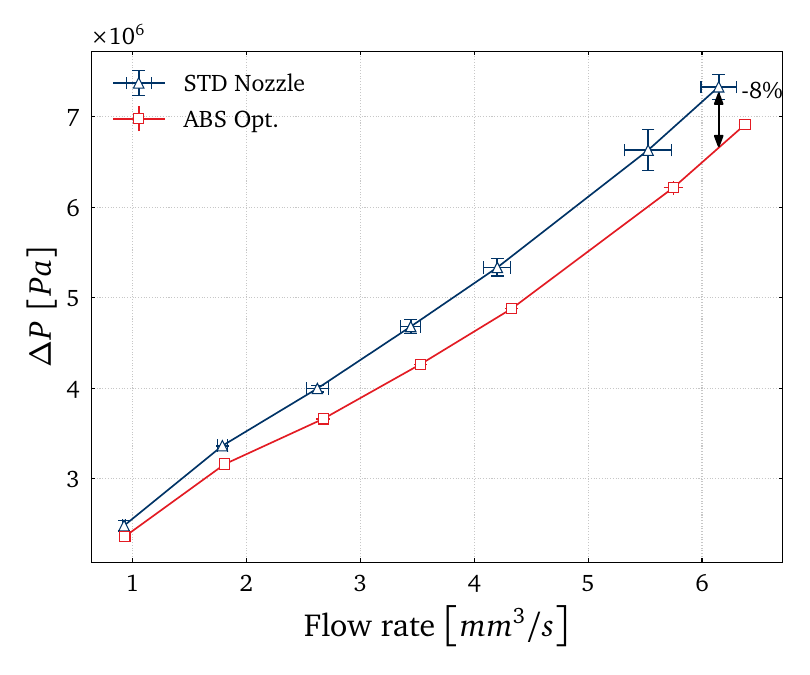}
    \caption{ABS.}
\end{subfigure}
\begin{subfigure}[t]{.45\textwidth}
    \centering
    \includegraphics[width=\textwidth]{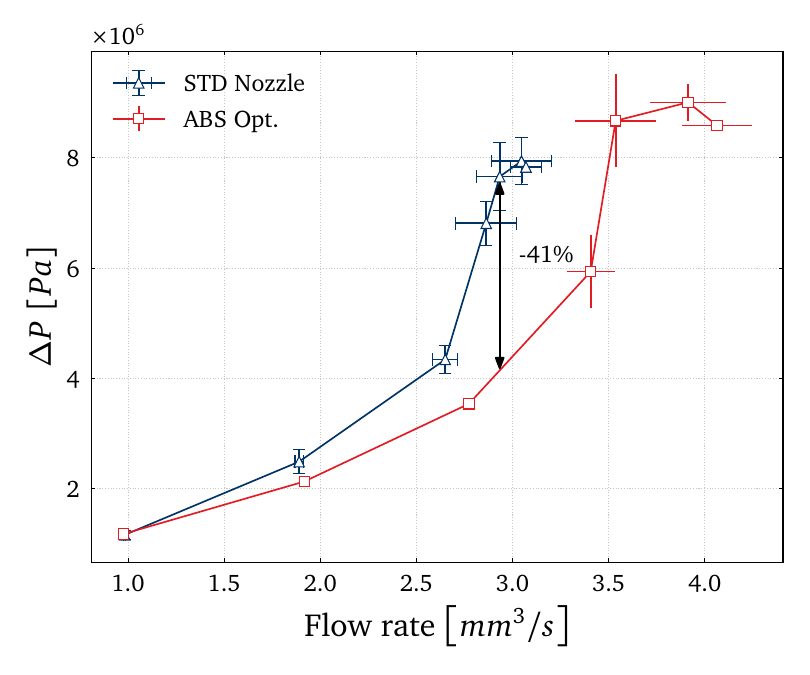}
    \caption{PET-CF.} 
\end{subfigure}
\begin{subfigure}[t]{.425\textwidth}
    \centering
    \includegraphics[width=\textwidth]{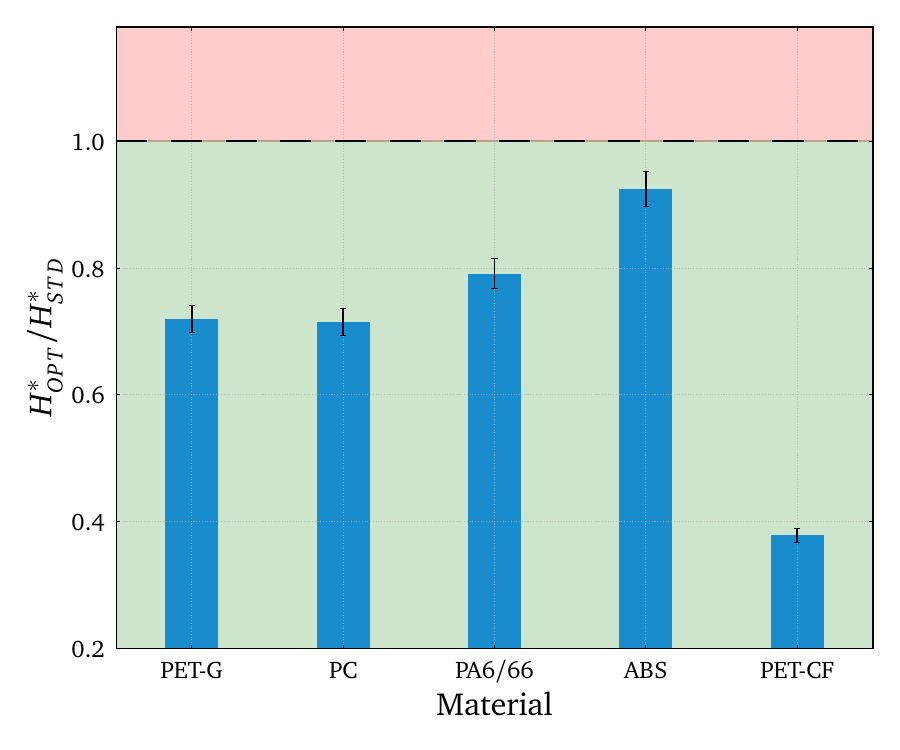}
    \caption{$H^*$ ratio.} 
\end{subfigure}
\caption{(a-e) Experimental pressure drop difference between the original standard core and the optimized geometries for each group of materials. (f) $H^*$ ratio between the standard and optimized shape per material at the maximum flow rate allowed for the standard nozzle (Table~\ref{tab:TotalPress_all_mat_geos}).}
\label{fig:nozzle_pressures}
\end{figure}

Figure~\ref{fig:nozzle_pressures} presents the pressure drop difference between the original nozzle and the different optimized nozzles for each polymer group~\cite{paper2}. It is possible to observe the improvement in pressure drop across all materials, confirming the numerical results. The experimental pressure drops are higher than the simulated ones, as already reported in our previous work \cite{sietsepaper}. This is because the numerical model exclusively accounts for the pressure drop in the nozzle ($\Delta P_{n}$), and it does not account for the extra pressure drop occurring in the annular gap due to the backflow. Therefore, the percentage of improvement in the pressure drop measured in the experiments is regarding the total pressure drop ($\Delta P_{total}$), sensed by the installed load cell (Fig. \ref{fig:monostrudel}), and therefore is not as great as in the numerical simulations, being muffled by the extra pressure drop occurring in the liquefier.

\noindent We found that the percentage of improvement in the experimental pressure drop ($\Delta P_{total}$ in Figure \ref{fig:nozzle_pressures}) is less significant than in the nozzle-specific pressure drop ($\Delta P_n$ in Table \ref{tab:drop_all_mat_geos}). This discrepancy arises because the contribution of the backflow zone to the total pressure drop ($\Delta P_{liq}=\Delta P_{total} - \Delta P_n$) carries more weight than the nozzle itself \cite{schuller_add_ma,sietsepaper}.

However, it is crucial to note that the pressure drop in the backflow zone is closely tied to its height ($H^*$ - see Equation \ref{eq:experimental}), which, in turn, depends on the pressure drop in the nozzle. By optimizing the nozzle shape to reduce pressure drop, we indirectly contribute to a lower pressure drop in the backflow region, making the optimized design less prone to backflow issues than the standard nozzle geometry.

\noindent Mathematically, this relationship is represented by the formula:

\begin{equation}
    H^*_{opt}=H^*_{std}\cdot \frac{\Delta P_{liq-opt}}{\Delta P_{liq-std}}.
\end{equation} 

As $\Delta P_{liq-opt}<\Delta P_{liq-std}$, it follows that $\Delta P_{liq-opt}/\Delta P_{liq-std}<1$, and consequently, $H^*_{opt} <H^*_{std}$ (Figure~\ref{fig:nozzle_pressures} f).

Furthermore, we observed an overshoot in the total pressure drop starting at 2.5 mm$^3$/s, a phenomenon previously noted with PLA and a similar geometry \cite{sietsepaper}. This study observed it in PA6/66 and PET-CF with the standard geometry, albeit for slightly higher flow rates. This behavior is likely due to the absence of a step in the tapered region of the standard nozzle compared to the other nozzle. Additionally, we found that feeders tend to slip due to the high total pressure drop for these two polymers.
Looking at the $\Psi_{10}\approx\eta \cdot \lambda$ values, PA6/66 ($\sim 670$ Pa$\cdot $s$^2$) and PET-CF ($\sim 28200$ Pa$\cdot$s$^2$) exhibit larger values than PLA, that has a value of $\sim450$ Pa$\cdot s^2$, and that explains that the onset of the elastic turbulence with the standard nozzle geometry happens at a similar flow rate. Nevertheless, it is surprising that ABS, which has a $\Psi_{10}\sim 582000$ Pa$\cdot$s$^2$, did not exhibit that behavior, although this may be due to the very marked shear thinning shown by ABS \cite{paper2}.

Reducing the total pressure with the optimized geometries also enables the possibility of achieving higher flow rates, especially for PA6/66 and PET-CF. Moreover, the optimized shape reduces vortex size, leading to more stable pressure drop readings. While not explored in this study, the reduction in recirculation size may have positive implications for nozzle clogging prevention, which has repercussions on printing performance and reliability.

\begin{figure}[htbp]
\centering
\includegraphics[width=0.6\linewidth]{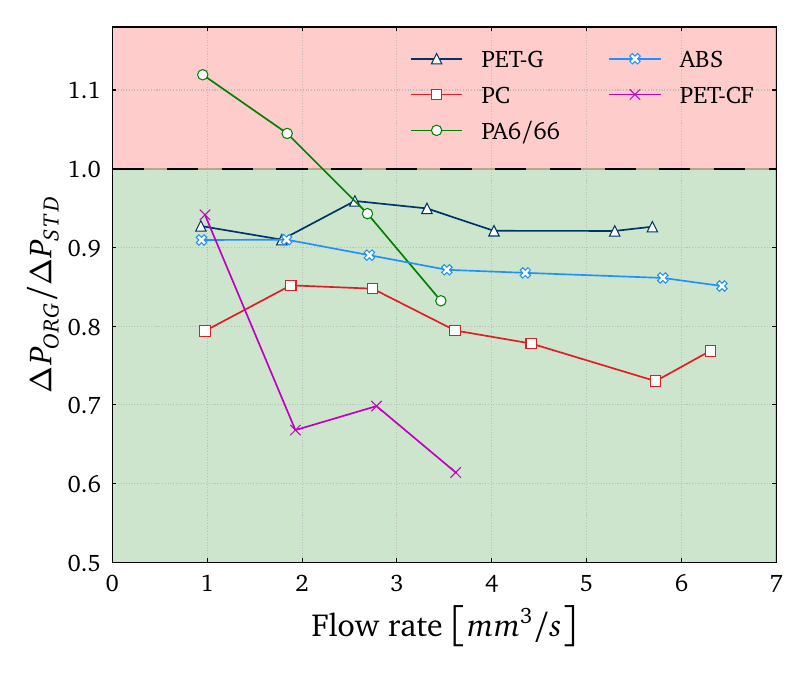}
\caption{Relative pressure drop between the organic and the standard nozzle geometry.}
\label{fig:org_comp}
\end{figure}

Despite the organic optimal geometry being optimized for PC at an extrusion velocity of 110~mm/s, the numerical results shown in Table~\ref{tab:drop_all_mat_geos} predicted that this geometry could also improve the nozzle pressure drop for all the materials. The experimental results for the organic shape (Figure \ref{fig:org_comp}) demonstrated a reduction in the total pressure drop not only for the polymers with similar rheological properties (PA6/66 being also from Group 2) but also for all the others. Due to feeder slip, there is a lack of measurements above $\sim$3.5~mm$^3$/s for PA6/66 and PET-CF when working with the standard geometry; therefore, the comparison was shown only until these flow rate values. The organic geometry only showed poorer results for PA6/66 at low flow rates, i.e. below 2~mm$^3$/s.

\begin{figure}[htbp]
\centering
\includegraphics[width=0.6\linewidth]{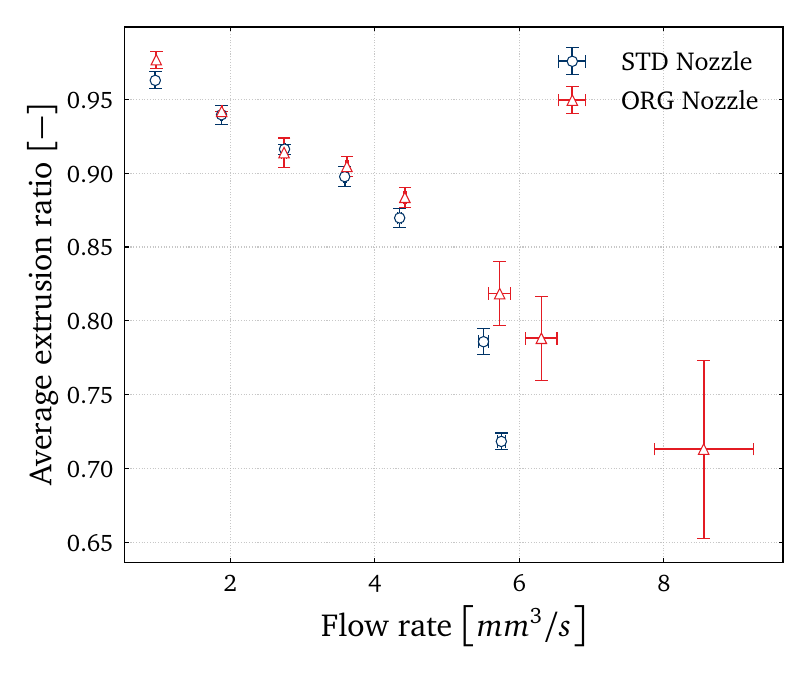}
\caption{Comparative between the performance of the original standard nozzle and the organic nozzle when operating with PC at different imposed flow rates. The experimental data set shows the average extrusion ratio as a function of the imposed flow rate, showing the feeder slip occurring in the standard nozzle above 5~mm$^3$/s.}
\label{fig:slip_pc}
\end{figure}

Reducing the recirculation size in the optimized geometries (present in Table \ref{tab:recirc_all_mat_geos}) also brings the benefits of minimizing the risk of nozzle clogging and material degradation in the nozzle. 
Reducing pressure drop is crucial for increasing printing speed while preventing feeder slip. As printing speed increases, extrusion speed must also rise, causing an increase in pressure drop. Higher pressure demands more force from the feeder, but excessive force may result in feeder slip. The feeder struggles to meet the required flow rate at high forces, resulting in under-extrusion and potential printing defects.
Figure \ref{fig:slip_pc} shows the feeder slip for PC in the standard and organic geometries. The average extrusion ratio is the ratio between the material flow rate sensed by the rotary encoder and the feeder rotation. It becomes evident that the reduced pressure on the organic geometry mitigates the possibility of having feeder slip issues, especially at higher flow rates.

\FloatBarrier
\subsection{Printing performance}
\label{sec:perform}

\begin{figure}[htbp]
	\begin{center}		
		\includegraphics[width=\textwidth]{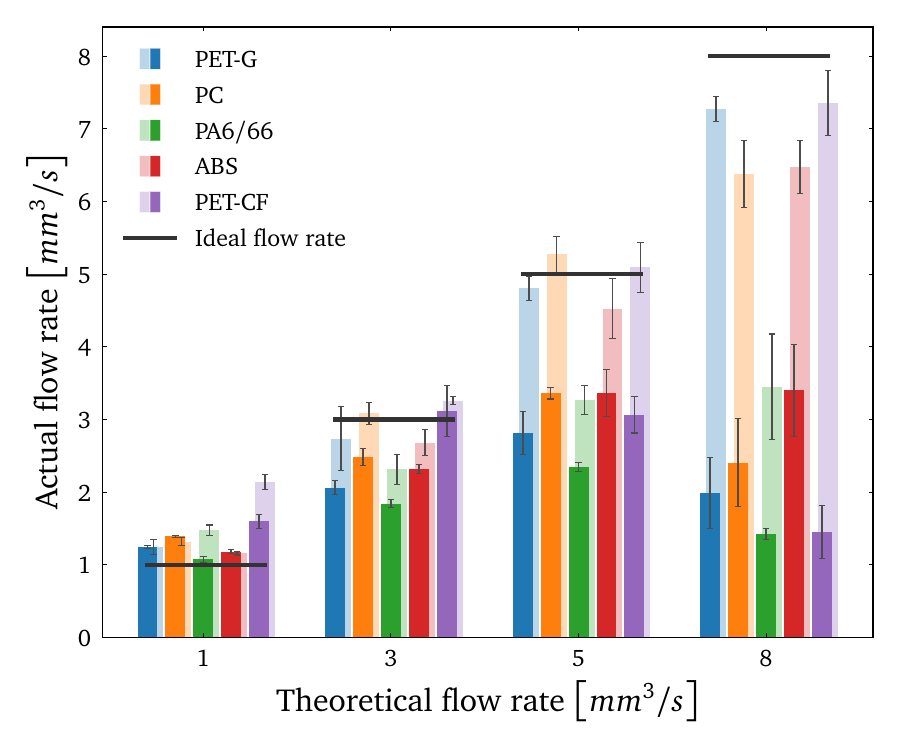}
		\caption{Actual \textit{vs} theoretical flow rate, comparing original (dark) and optimized (light) nozzles.}
		\label{fig:oct_barplot}
	\end{center}
\end{figure}

Figure \ref{fig:zigzags} presents the print-bed after a 1-layer test print, where it is possible to see the difference in line width along the zigzag pattern. These prints were then submitted to an optical tomography scan to evaluate the lines' dimensions and calculate the actual polymer volumetric flow rate, computed from the width and height measured with OCT ($w\cdot h\,\cdot$ print speed). Figure \ref{fig:oct_barplot} shows how the optimized geometries allows for depositing the amount of material pre-selected in the custom script. Therefore, it is confirmed that the optimized geometries (light colors) enabled a more accurate extrusion process (less feeder slip) resulting in a more accurate size of the deposited lines. Moreover, these new nozzles would also increase the productivity of the FFF process.






\FloatBarrier

\section{Conclusions and remarks}

In conclusion, our study not only contributes to Fused Filament Fabrication (FFF) nozzle design to address crucial challenges in additive manufacturing, such as pressure-related issues and material deposition precision, but also seeks to bridge the gap between numerical optimization and practical 3D printing performance improvements. By focusing on nozzle geometry and employing Computational Fluid Dynamics (CFD) simulations, our research offers insights into how optimized nozzle designs can significantly enhance material extrusion, improve printing efficiency, and ensure the overall quality and reliability of 3D-printed objects.

In this work, we demonstrated that by optimizing nozzle shape, we can effectively control the flow patterns. The optimized shape proved less prone to elastic instabilities and exhibited a Newtonian-like behavior, in the sense of upstream vortices suppression.
The objective function, minimization of pressure drop, depends on viscous dissipation, shear- ($N_{1-S}$) and extension-induced normal stresses ($N_{1-E}$). The optimal shape solution promotes either shear or extensional flow depending on the material's rheological properties; whereas polymers belonging to Groups 1 and 2 benefited from a promoted shear flow, polymers included in Group 3, having higher polymer viscosity, a promotion in the extensional flow provided lower pressure drop.
We revealed that pressure drop in the nozzle zone significantly impacts total pressure drop and equilibrium height ($H^*$). By optimizing nozzle shape and indirectly reducing backflow-related pressure drop, we make the design less prone to backflow issues.
Nozzle optimization enhances flow rate control and opens the possibility for higher printing speeds and productivity. Reduced recirculation size, minimized feeder slip, and improved line dimensions positively impact printing quality and reliability.

The authors demonstrate a novel method to design and fabricate custom nozzles for fused filament fabrication (FFF) using metal 3D printing. This work has both industrial relevance and scientific significance, as it opens new possibilities for enhancing 3D printing performance and quality. Furthermore, in the context of the emerging Artificial Intelligence (AI) revolution, this work could lead to a new paradigm in additive manufacturing, where an AI system can control an FFF factory with a metal 3D printer that can produce optimal nozzles for different materials. The key requirement for this scenario is a comprehensive rheological database of various polymers under different rheometric flows and temperature conditions, enabling the AI system to select the appropriate constitutive model, parameters and objective function for each material and design the best nozzle accordingly.

\FloatBarrier

\section*{Acknowledgements}
The authors would like to thank Francisco Vide Coelho de Almeida, Johan Versteegh for the fruitful discussion and selfless support. FJGR and TS acknowledge the financial support from UltiMaker B.V., and also national funds through FCT/MCTES (PIDDAC) and projects LA/P/0045/2020 (ALiCE), UIDP/00532/2020 (CEFT), and the program Stimulus of Scientific Employment, Individual Support-2020.03203.CEECIND. TS acknowledges the support of the Erasmus+ programme of the European Union.

\section*{CRediT authorship contribution statement}

\noindent \textbf{Tom\'as Schuller}: Conceptualization, Data Curation, Formal analysis, Investigation, Methodology, Software, Validation, Visualization, Writing - Original Draft.\\
\textbf{Maziyar Jalaal}: Methodology, Supervision, Validation, Visualization, Writing - Review \& Editing.\\
\textbf{Paola Fanzio}: Conceptualization, Funding acquisition, Methodology, Resources, Supervision, Validation, Visualization, Writing - Review \& Editing.\\
\textbf{Francisco J. Galindo-Rosales}: Conceptualization, Funding acquisition, Methodology, Project administration, Resources, Supervision, Validation, Visualization, Writing - Review \& Editing.

\section*{Declaration of competing interest}

The authors declare that they have no known competing financial interests or personal relationships that could have appeared to influence the work reported in this paper.

\section*{Data avaliability}
All data generated or analysed during this study are included in this article and its supplementary information files.

\newpage
\appendix
\setcounter{figure}{0}
\FloatBarrier
\section[\appendixname~\thesection]{ - Governing equations}
\label{app:const_eq}

Assuming isothermal and steady flow through the extrusion nozzle in all materials and simulations, we neglected the energy equation. The molten polymer was treated as an incompressible fluid, hence,
 
 \begin{equation}
  \label{eq: Continuity}
  \nabla \cdot \boldsymbol{u} =0. 
\end{equation} 

\noindent The momentum conservation equation reads:
 
\begin{equation}
  \label{eq: Momentum}
  \rho \, (\boldsymbol{u}\cdot\nabla \boldsymbol{u}) =-\nabla P +\nabla\cdot\boldsymbol{\tau}.
\end{equation} 

\noindent In this scenario, we examined a steady-state flow ($\frac{\partial\boldsymbol{u}}{\partial t}=0$), with the negligible effects of gravity excluded. Here, $P$ represents pressure, $\rho$ stands for density, and the stress tensor $\boldsymbol{\tau}$ includes polymer contributions, as defined in the Giesekus model equations \cite{giesekus,GIESEKUS1985349} :

    \begin{equation}
        \boldsymbol{\tau}_s = \eta_s \left(\nabla\boldsymbol{u}+\nabla\boldsymbol{u}^T\right),
        \label{Eq:giesekus_s}
    \end{equation}

    \begin{equation}
        \boldsymbol{\tau} + \lambda \overset{\triangledown}{\boldsymbol{\tau}}+\alpha\frac{\lambda}{\eta_p}\left(\boldsymbol{\tau}\cdot\boldsymbol{\tau}\right) =\eta_p \left(\nabla\boldsymbol{u}+\nabla\boldsymbol{u}^T\right),     \label{Eq:giesekus}
    \end{equation}

\noindent where $\overset{\triangledown}{\boldsymbol{\tau}}$ denotes the upper convected derivative of the stress tensor. , $\eta_s$ indicates the solvent viscosity, $\lambda$ represents the relaxation time, $\alpha$ is the dimensionless mobility factor, $\eta_p$ indicates the polymer viscosity, and $\boldsymbol{u}$ represents the velocity field. Notably, $\alpha$ assumes a critical role in governing both the extensional viscosity and the ratio of the second normal stress difference to the first in this model.  To simulate viscoelastic fluid flows, it is a common approach to split the total extra-stress tensor in a solvent contribution ($\boldsymbol{\tau}_s$) and a polymeric contribution ($\boldsymbol{\tau}$), $\boldsymbol{\tau}'=\boldsymbol{\tau}+\boldsymbol{\tau}_s$ \cite{rheoTool}.

To address Eqs. \ref{eq: Continuity}, \ref{eq: Momentum}, \ref{Eq:giesekus_s}, and \ref{Eq:giesekus}, we employed the rheoTool library~\cite{rheoTool,pimenta1} for OpenFOAM\R. To ensure numerical stability when dealing with the viscoelastic fluid, we utilized both the log-conformation formulation of the constitutive equation and a high-resolution scheme (CUBISTA)~\cite{AlvesetalAnnRevFluMech2021}.

\FloatBarrier
\section[\appendixname~\thesection]{ - Grid convergence study}
\label{app:a}

\begin{figure}[htbp] 
  \begin{subfigure}[b]{0.5\linewidth}
    \centering
    \includegraphics[width=.9\linewidth]{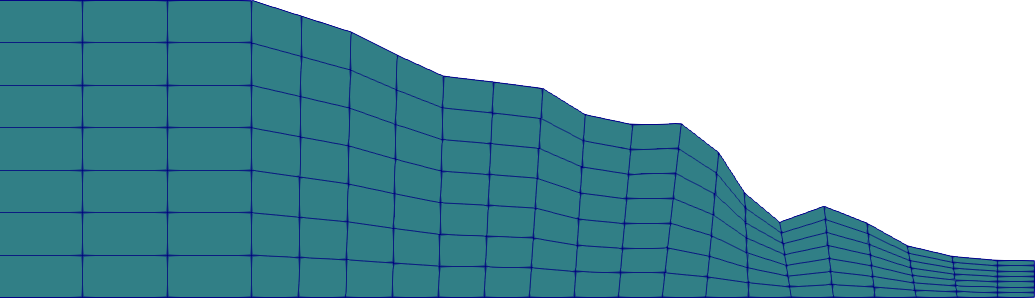} 
    \caption{Coarse mesh} 
        \vspace{4ex}
\end{subfigure}
  \begin{subfigure}[b]{0.5\linewidth}
    \centering
    \includegraphics[width=.9\linewidth]{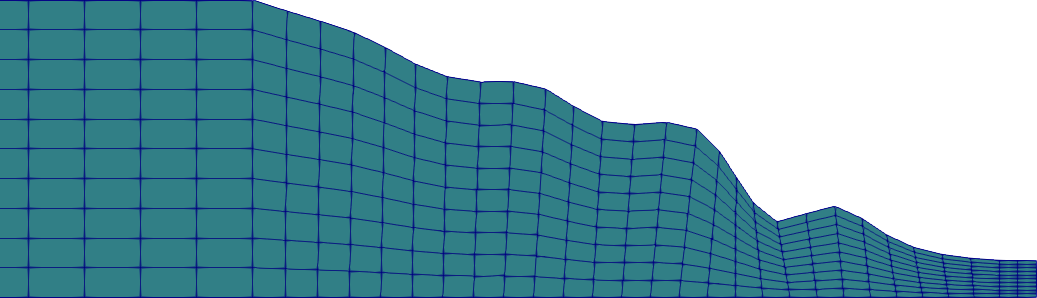} 
    \caption{Medium mesh} 
    \vspace{4ex}
\end{subfigure}\\
  \begin{subfigure}[b]{0.5\linewidth}
    \centering
    \includegraphics[width=.9\linewidth]{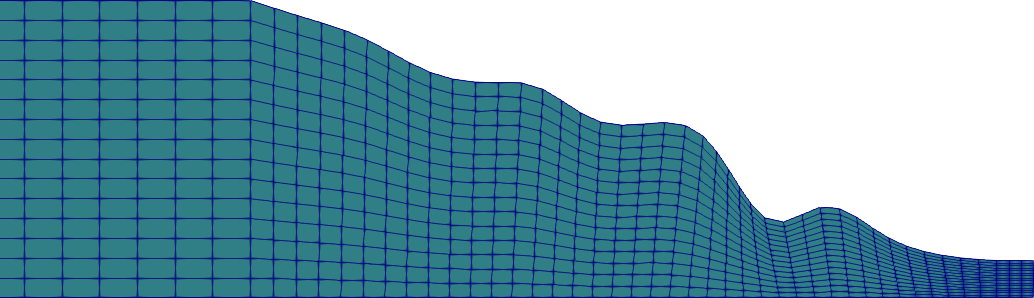} 
    \caption{Fine mesh} 
 \end{subfigure}
  \begin{subfigure}[b]{0.5\linewidth}
    \centering
    \includegraphics[width=.9\linewidth]{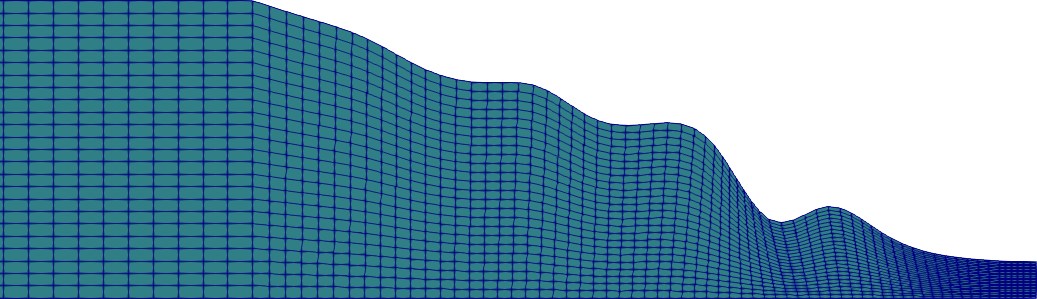} 
    \caption{Very fine mesh} 
\end{subfigure}
 \caption{Mesh sizes used in the GCS.}
 \label{fig:gcs_meshes}
\end{figure}

Using the gradient descent-generated geometry, a grid convergence study (GCS) was performed, with the results being shown in Table \ref{tab:gcs_tables} and the relative mesh sizes in Figure \ref{fig:gcs_meshes}. The study was performed following the protocol of the GCS by \cite{nasaGCS}, that is based on the works of P.J. Roache and L. F. Richardson \cite{Roache1972, Roache2006-dc,Roache1993,1911,1927}. 

\begin{table}[htbp]
    \centering
    \caption{Grid Convergence Study.}
    \label{tab:gcs_tables}
    \begin{subtable}{\linewidth}%
        \centering
        \caption{Initial Parameters.}
        \label{tab:gcs_table_1}
        \begin{tabular}{ccc}
            \toprule
            Grid \# & Relative Grid Size & Pressure drop [Pa] \\ \midrule
            1 & 2.25 & 261580.88 \\
            2 & 1.5 & 259889.79 \\
            3 & 1 & 309247.51 \\
            4 & 0.666 & 309269.27 \\ 
            \bottomrule
        \end{tabular}
    \end{subtable}%
    \\
    \begin{subtable}{\linewidth}
        \centering
        \caption{Study Results.}
        \label{tab:gcs_table_2}
        \begin{tabular}{cp{3.4cm}p{3cm}p{3cm}c}
            \toprule
            Grids & Approx. Error between geometries & Extrap. Error to next refinement & Extrap. refined result & GCI      \\ \midrule
            4 3 2 & 0.15960       & 0.005631           & 310998.60  & 0.0071 \\
            3 2 1 & 0.00007       & 0.000000           & 309269.28  & 0.0000 \\
            \bottomrule
        \end{tabular}
    \end{subtable}
\end{table}

\FloatBarrier
\section[\appendixname~\thesection]{ - Supplementary optimization information}
\label{app:b}

The variables of the SOGA optimization are presented in Table \ref{tab:soga_table_all}.

\begin{table}[htbp]
\centering
\caption{SOGA parameters.}
\resizebox{0.6\textwidth}{!}{%
\begin{tabular}{lc}
\toprule Parameter & Value \\ \midrule
Maximum \# of evaluations & 1500 \\
Random seed & 10983 \\
Population Size & 10 \\
Initialization type & Unique random \\
Mutation type & Replace uniform \\
Mutation rate & 0.15   \\
Crossover type & Shuffle random \\
\# Offspring & 4 \\
\# Parents & 4 \\
Crossover rate & 0.8 \\
Replacement type & Elitist  \\
Fitness type  & Merit function \\
Constraint penalty  & 0.9 \\
Convergence type & Average fitness tracker \\
\# Generations & 200   \\
Percent change & 0.05 \\
\bottomrule
\end{tabular}}
\label{tab:soga_table_all}
\end{table}

DAKOTA\C allows for constrained optimization runs for simple cases, but struggles to obey these constraints when the problem becomes more complex. To solve this issue, a cleaning script was developed to not only select the individuals from each population that followed these preset rules, but also to begin with a predefined quasi-random population that complies with the rule-set \cite{foamscripts}.

\FloatBarrier
\section[\appendixname~\thesection]{- Additional numerical and experimental results}
\label{app:e}

\begin{figure}[htbp]
\centering
\begin{subfigure}{.29\textwidth}
    \centering
    \includegraphics[width=\textwidth]{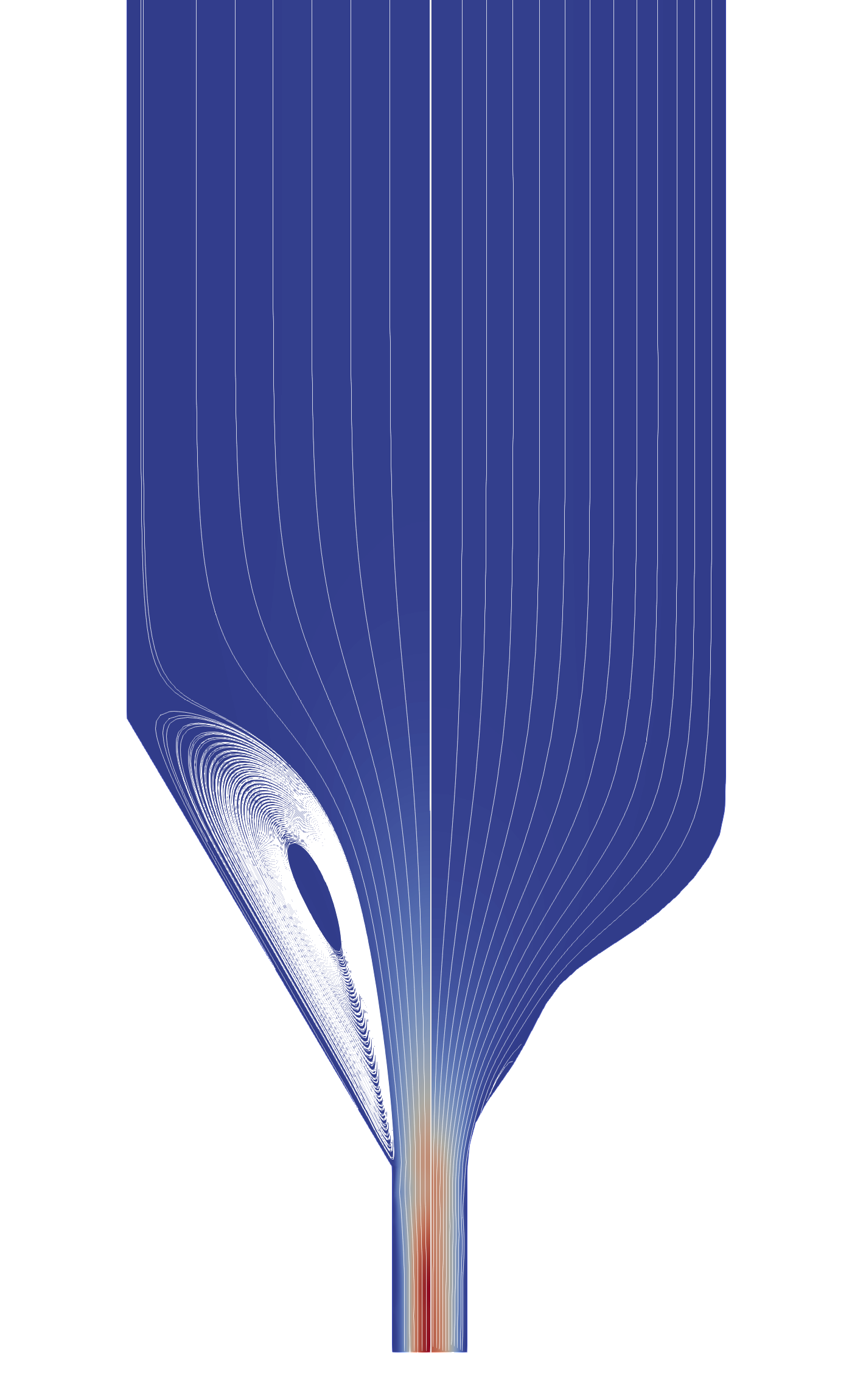}
    \caption{PET-G.}
\end{subfigure}%
\begin{subfigure}{.29\textwidth}
    \centering
    \includegraphics[width=\textwidth]{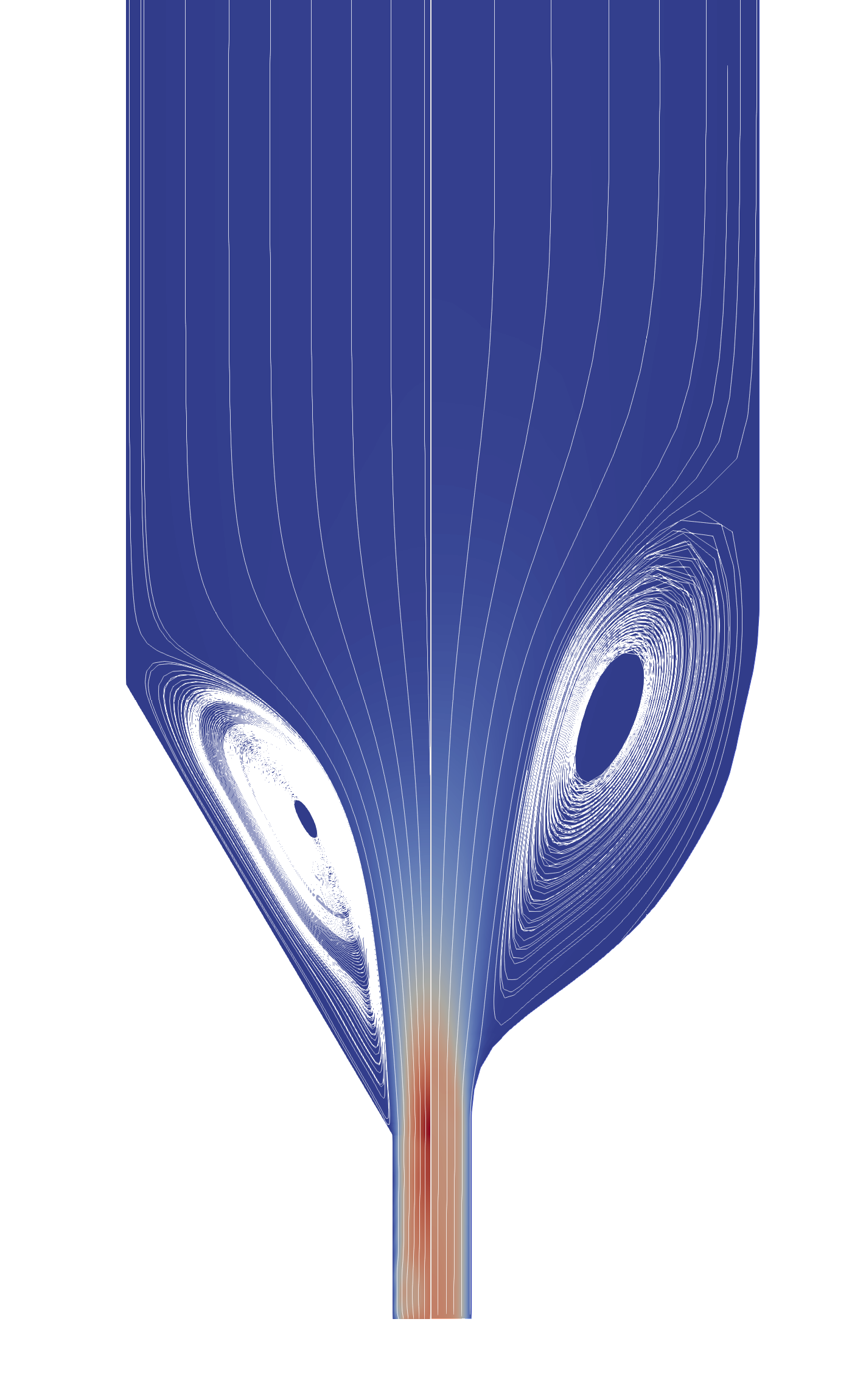}
    \caption{PC.}
\end{subfigure}
\begin{subfigure}{.29\textwidth}
    \centering
    \includegraphics[width=\textwidth]{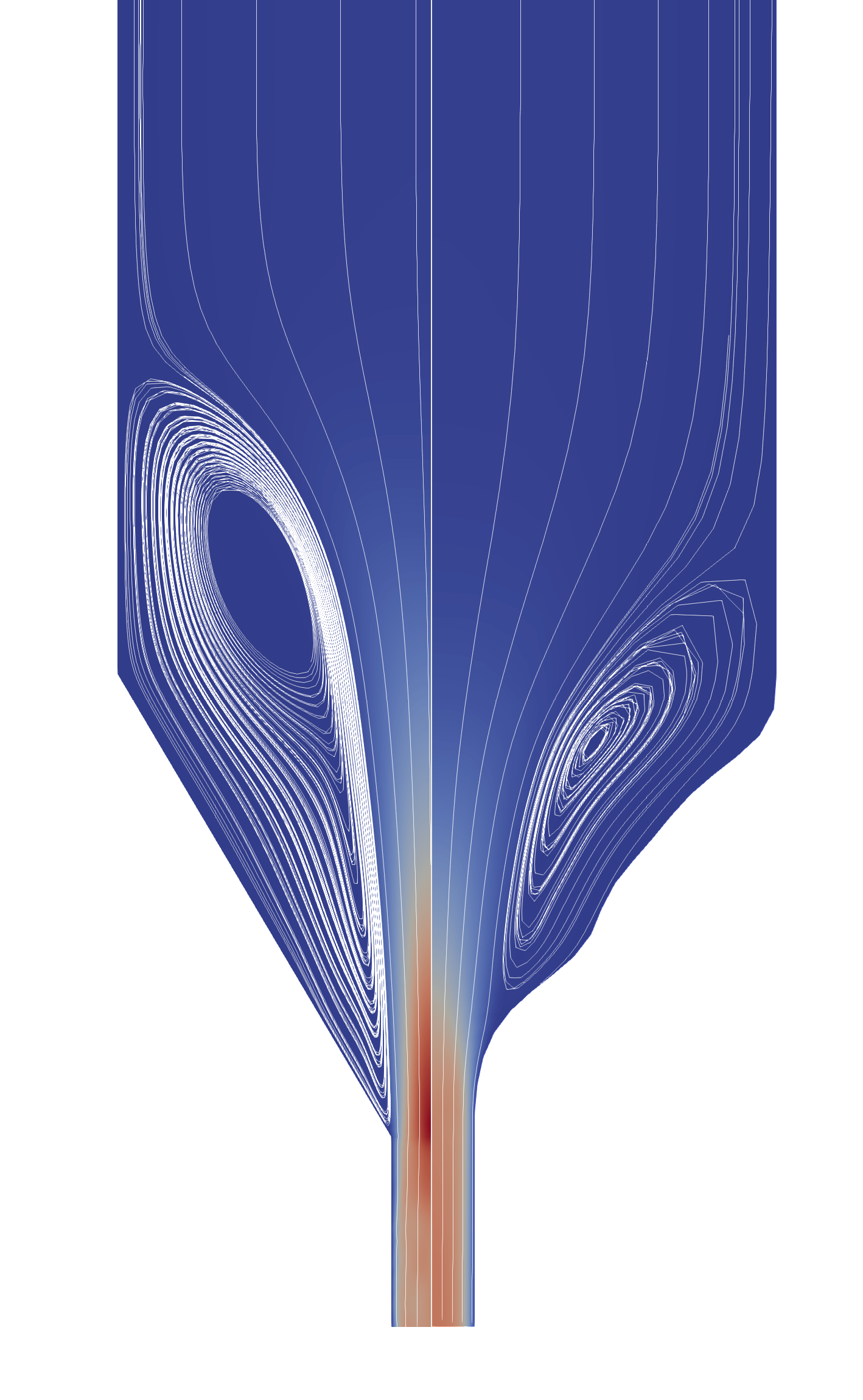}
    \caption{PA6/66.}
\end{subfigure}
\begin{subfigure}{.29\textwidth}
    \centering
    \includegraphics[width=\textwidth]{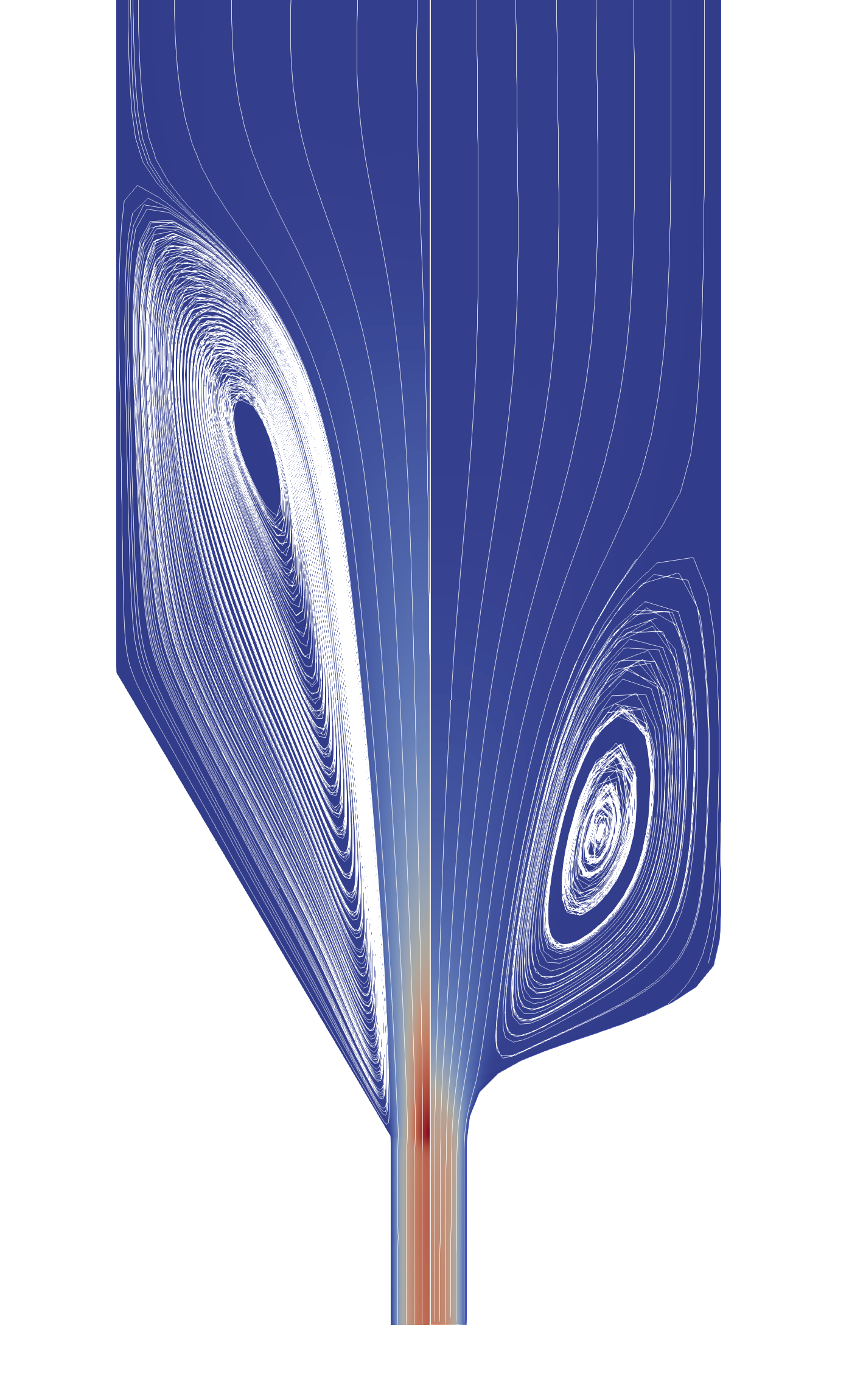}
    \caption{ABS.}
\end{subfigure}
\begin{subfigure}{.299\textwidth}
    \centering
    \includegraphics[width=\textwidth]{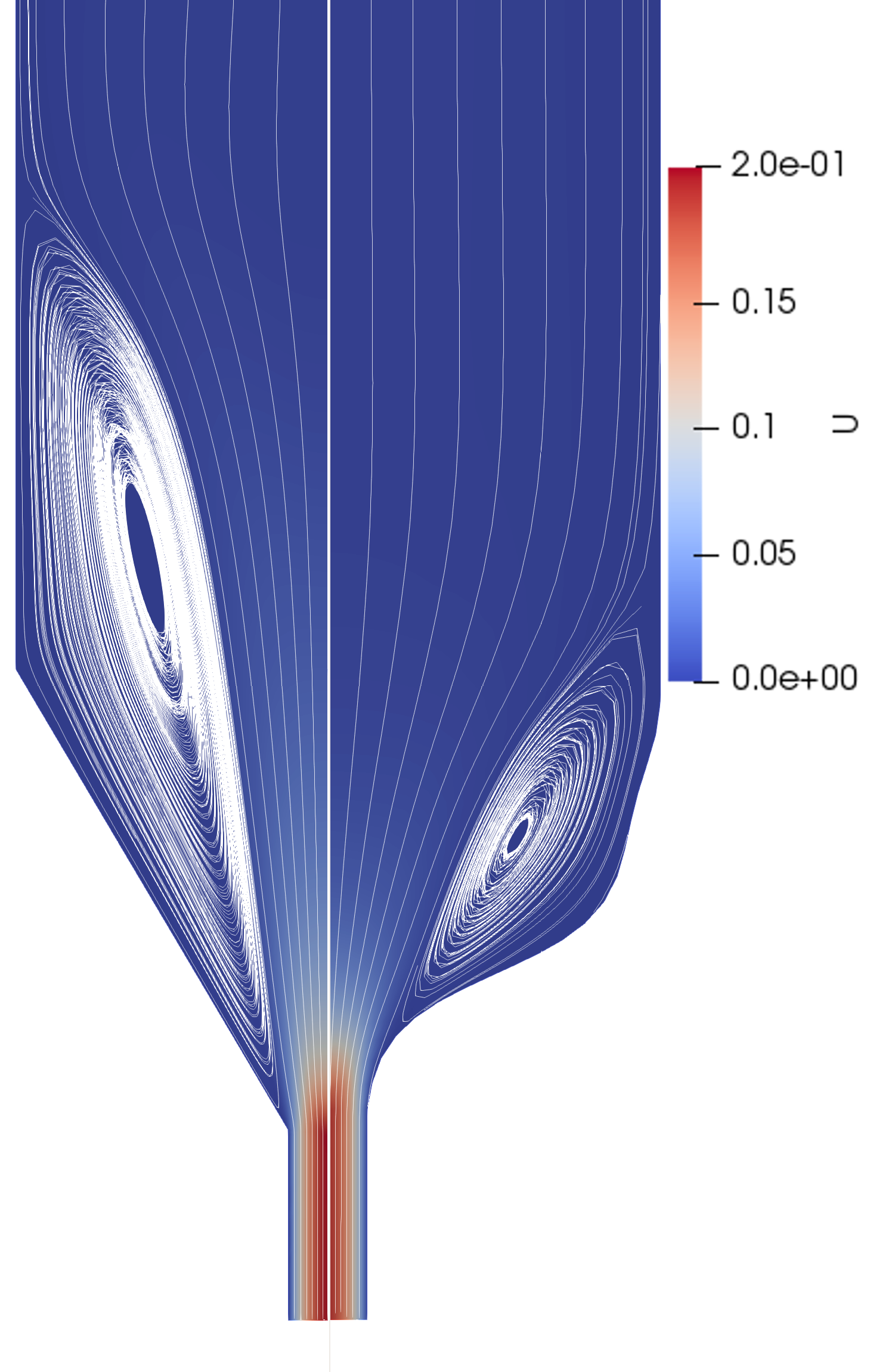}
    \caption{PET-CF.}
\end{subfigure}
\caption{Comparison between standard and optimized geometries regarding the velocity field (in mm/s) and streamline per material at 110 mm/s.}
\label{fig:streamlines}
\end{figure}

\begin{figure}[htbp]
\centering
\begin{subfigure}{.29\textwidth}
    \centering
    \includegraphics[width=\textwidth]{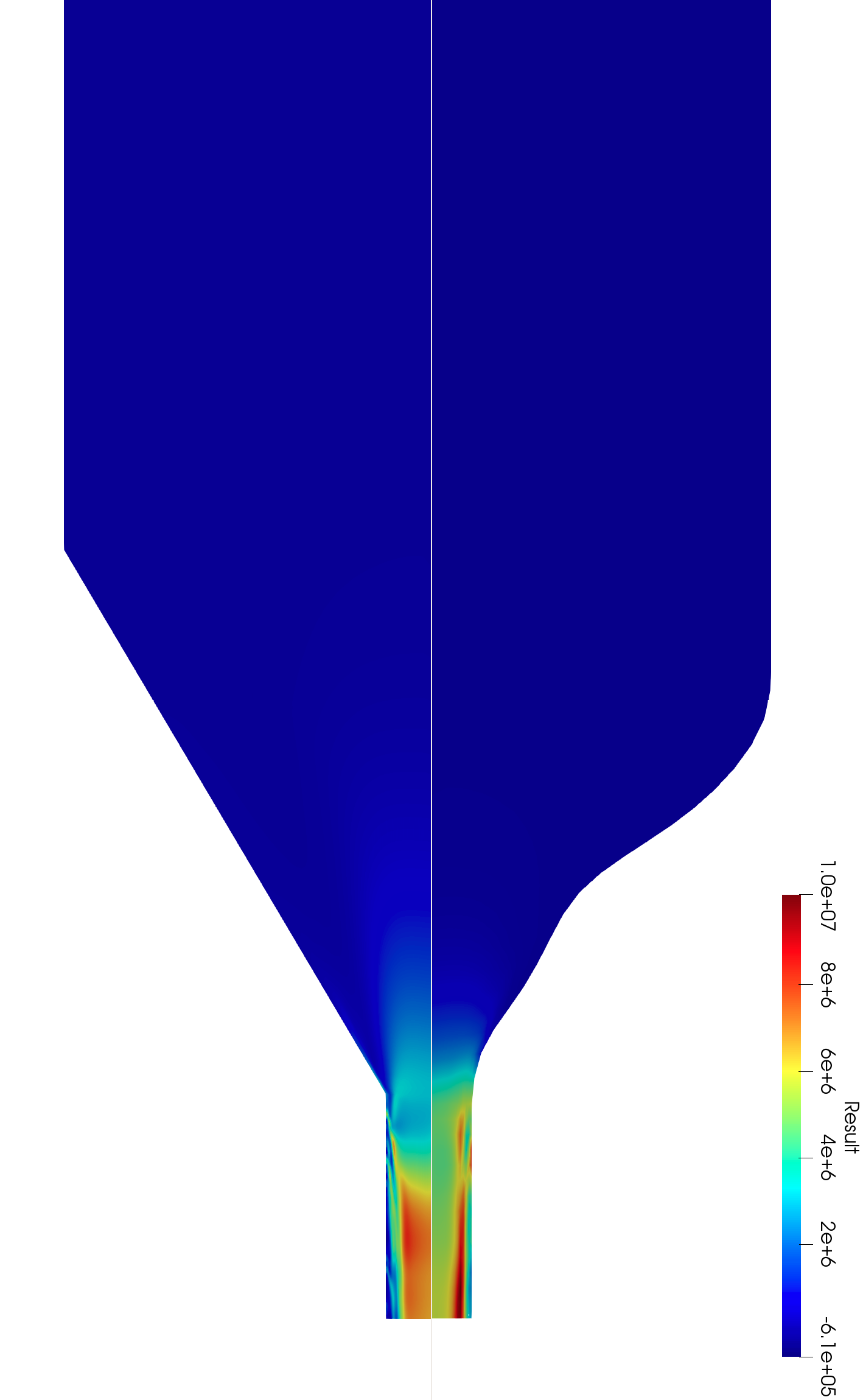}
    \caption{PET-G.}
\end{subfigure}%
\begin{subfigure}{.29\textwidth}
    \centering
    \includegraphics[width=\textwidth]{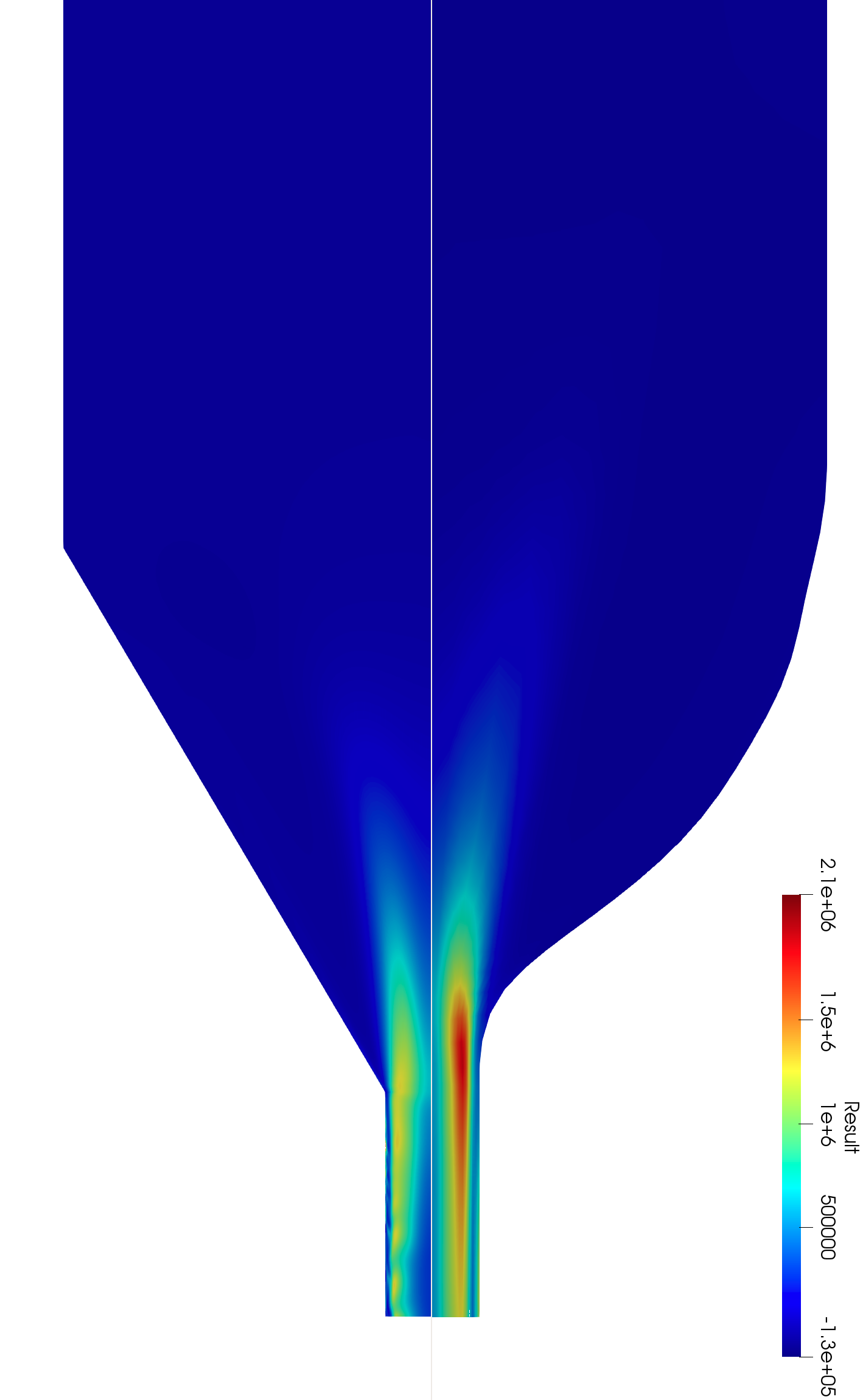}
    \caption{PC.}
\end{subfigure}
\begin{subfigure}{.29\textwidth}
    \centering
    \includegraphics[width=\textwidth]{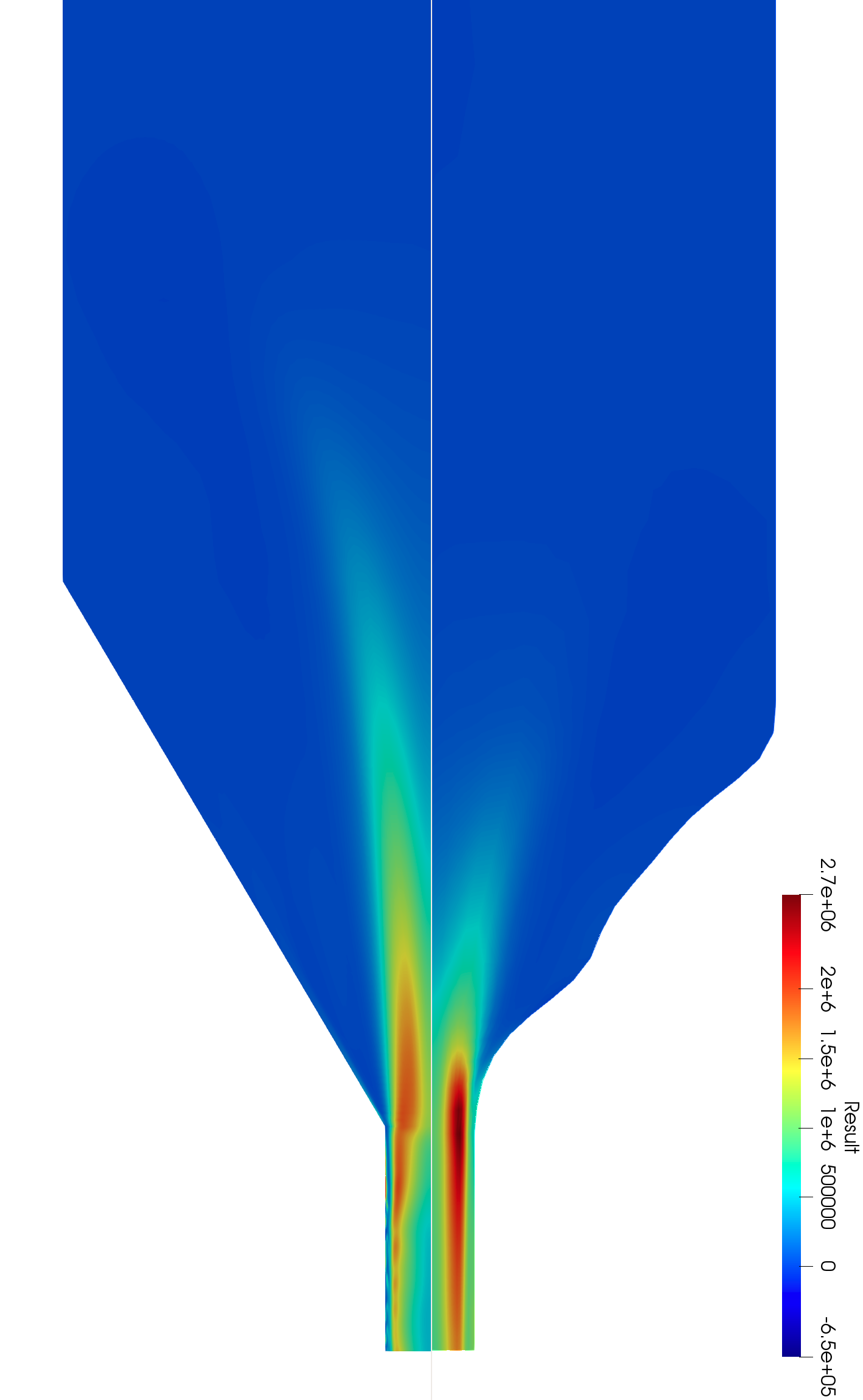}
    \caption{PA6/66.}
\end{subfigure}
\begin{subfigure}{0.29\textwidth}
    \centering
    \includegraphics[width=\textwidth]{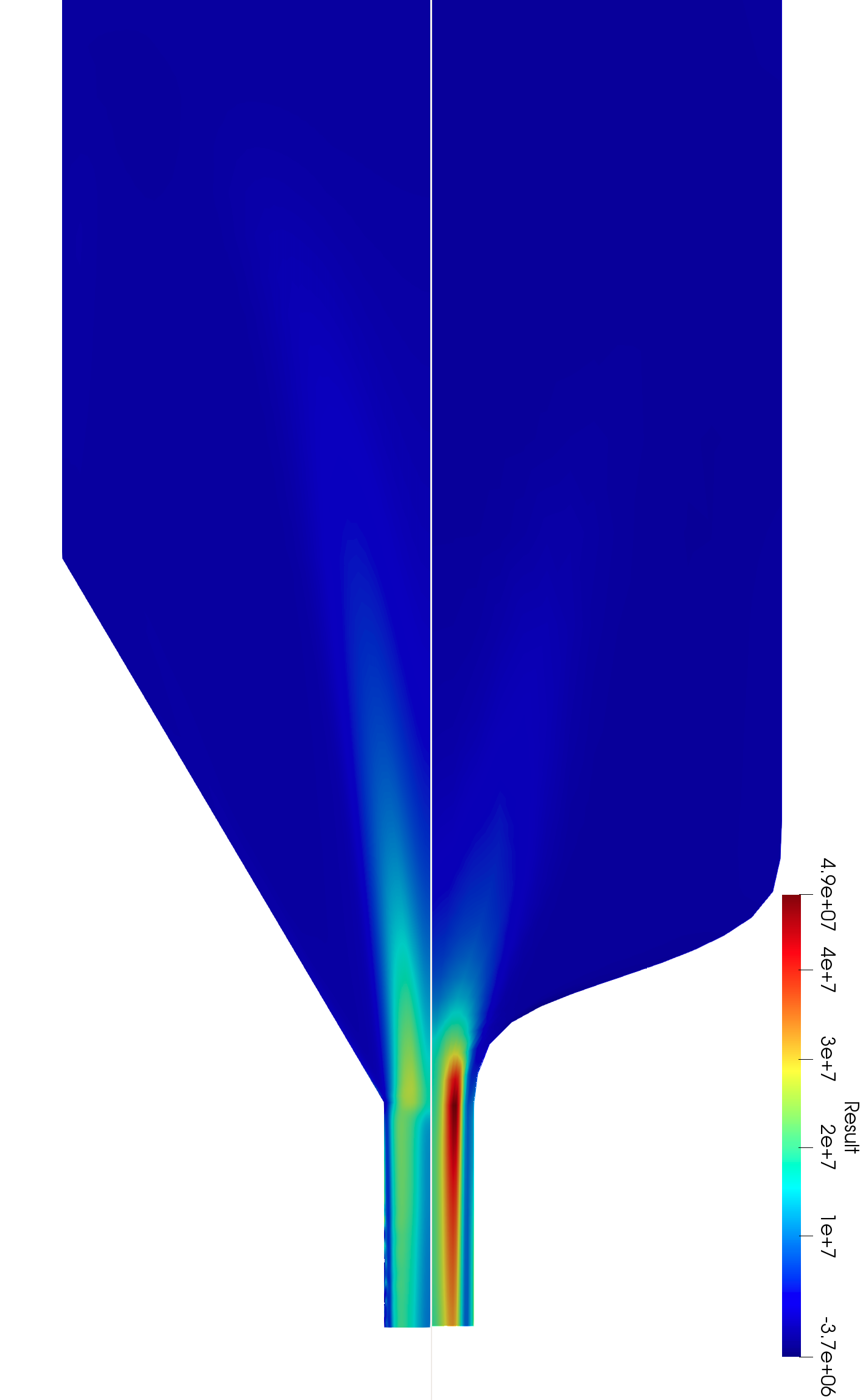}
    \caption{ABS.}
\end{subfigure}
\begin{subfigure}{.29\textwidth}
    \centering
    \includegraphics[width=\textwidth]{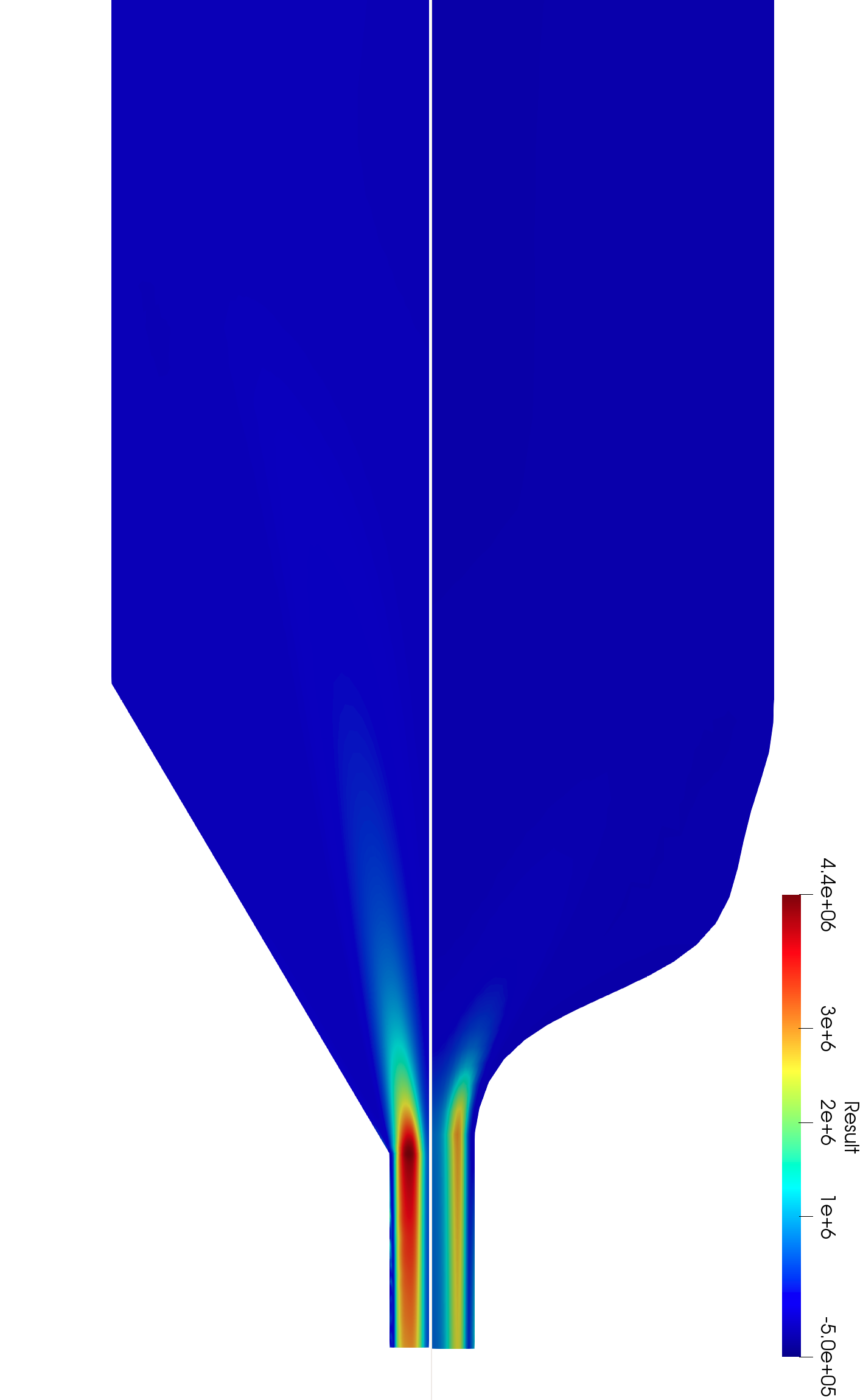}
    \caption{PET-CF.}
\end{subfigure}
\caption{Comparison between standard and optimized geometries regarding normal stress difference  (in Pa) per material at 110 mm/s.}
\label{fig:taus}
\end{figure}

Figures~\ref{fig:streamlines} and \ref{fig:taus} compares the flow-type, velocity field and streamlines, and $\Delta \tau$, respectively, between the original nozzle and the optimal shape for the different materials at extrusion velocity of 110~mm/s. It is possible to see that the optimized geometries generate a wider central path, with the vortexes being confined further from the center. 

Table \ref{tab:drop_all_mat_geos} presents the pressure drop improvement in percentage in relation to the standard geometry for all materials. Each column is for an optimal geometry obtained for a certain material at an extrusion velocity of 110~mm/s, whereas the rows in the table represent the material flowing through each geometry. Table \ref{tab:recirc_all_mat_geos} details the upstream vortex reductions in relation to the standard geometry for all materials. The highest improvement percentages for each material (constrained) are highlighted in bold. Moreover, a dark green background highlights the best average improvement, whereas a dark red is given for the worst average improvement. In Table \ref{tab:drop_all_mat_geos}, the organic geometry results are also present (italic), presenting a very high performance across all materials.

\begin{table}[htbp]
\caption{Numerical improvement of the nozzle pressure drop in relation to the standard geometry at an extrusion velocity of 110 mm/s. }
\label{tab:drop_all_mat_geos}
\resizebox{\textwidth}{!}{%
\begin{tabular}{lccccccc}
\toprule   & \textbf{GEO PA6-66} & \textbf{GEO PC}  & \textbf{GEO ORG-PC}   & \textbf{GEO ABS} & \textbf{GEO PET-CF} & \textbf{GEO PET-G}  & \textbf{AVG}        \\
\textbf{PA6-66} & \textbf{69.80\%} & 58.80\% &\textit{ 78.80\%} & 50.60\% & 54.67\% & 58.50\%  & \cellcolor[HTML]{009901}\textbf{61.86\%} \\
\textbf{PC}     & 24.77\% & 17.23\% &\textit{ 66.77\%} & \textbf{25.42\%} & 16.31\% & 18.46\%  & \cellcolor[HTML]{F8FF00}\textbf{28.16\%} \\
\textbf{ABS}    & 29.37\% & 28.05\% & \textit{73.60\%} & \textbf{47.92\%} & 27.46\% & 27.85\%  & \cellcolor[HTML]{32CB00}\textbf{39.04\%} \\
\textbf{PET-CF} & -7.68\% & 0.49\% & \textit{66.08\%} & 0.03\%  & \textbf{4.65\%}  & -3.65\%  & \cellcolor[HTML]{CB0000}\textbf{9.99\%} \\
\textbf{PET-G}  & 31.22\% & 28.49\%  &\textit{ 67.63\%} & 2.16\%  & -20.86\% & \textbf{65.47\%} & \cellcolor[HTML]{F8FF00}\textbf{29.02\%} \\
\textbf{AVG}    & \cellcolor[HTML]{F8FF00}\textbf{29.50\%} & \cellcolor[HTML]{F8FF00}\textbf{26.61\%}  & \cellcolor[HTML]{009901}\textbf{\textit{70.57\%}} & \cellcolor[HTML]{F8FF00}\textbf{25.22\%} & \cellcolor[HTML]{FE0000}\textbf{16.45\%} & \cellcolor[HTML]{32CB00}\textbf{33.33\%}  &         \\\bottomrule 
\end{tabular}%
}
\end{table}

\begin{table}[htbp]
\caption{Numerical improvement of the recirculation size in relation to the standard geometry at an extrusion velocity of 110 mm/s.}
\label{tab:recirc_all_mat_geos}
\resizebox{\textwidth}{!}{%
\begin{tabular}{lcccccc}
\toprule & \textbf{GEO PA6-66} & \textbf{GEO PC}  & \textbf{GEO ORG-PC}  & \textbf{GEO ABS} & \textbf{GEO PET-CF} & \textbf{GEO PET-G} \\\midrule
\textbf{PA6-66} & 46.86\% & 37.43\%     & \textit{70.36\%}    & \textbf{50.60\% }         & 42.94\% & 37.96\%            \\
\textbf{PC}     & 4.04\%  & -8.28\%      & \textit{11.42\%}      & \textbf{14.78\%}          & 2.02\%  & 0.74\% \\
\textbf{ABS}    & 29.48\% & 27.82\%      & \textit{37.68\%}      & \textbf{40.10\%}          & 33.33\% & 27.38\%            \\
\textbf{PET-CF} & 82.78\% & 40.13\%     & \textit{11.75\%}       & 45.91\%          & 49.75\% & \textbf{98.42\%}            \\
\textbf{PET-G}  & 88.73\% & 51.30\%      & \textit{75.07\%}      & 49.33\%          & 41.48\% & \textbf{100.00\%}     \\\bottomrule
\end{tabular}%
}
\end{table}

\begin{table}[htbp]
\centering
\caption{Experimental values of the total pressure drop in MPa for each material and each nozzle geometry. Presented flow rate values are the upper extrusion limit of standard geometry.}
\label{tab:TotalPress_all_mat_geos}
\resizebox{0.8\textwidth}{!}{%
\begin{tabular}{lcccc}
\toprule Material/Geometry & \textbf{STD} & \textbf{PET-G}  & \textbf{ORG}  & \textbf{ABS}  \\\midrule
\textbf{PA6-66} (@3.4mm$^3$/s) & 9.01 & 7.64 & 6.84 & 6.35 \\
\textbf{PC} (@5.7mm$^3$/s) & 10.9  & 8.54 & 7.48 & 8.40 \\
\textbf{ABS} (@6.1mm$^3$/s) & 7.33 & 6.32 & 6.91 & 6.23 \\
\textbf{PET-CF} (@3mm$^3$/s) & 7.84 & 3.09 & 3.03 & 3.54 \\
\textbf{PET-G} (@5mm$^3$/s) & 9.66 & 7.97 & 8.34 & 7.99 \\\bottomrule
\end{tabular}%
}
\end{table}

\FloatBarrier
\section[\appendixname~\thesection]{ - Metal binder jetting nozzles}
\label{app:c}

Metal binder jetting is an advanced additive manufacturing technique that is used for the production of intricate metal parts. This process combines the precision of 3D printing with the strength and durability of metal materials. Using a layer-by-layer approach, metal binder jetting involves depositing fine metal powder and binding agents to create complex and highly detailed components. After printing, the parts are sintered in a high-temperature furnace, fusing the metal particles together to form a solid, fully dense metal object.

\begin{figure}[htbp]
\centering
\begin{subfigure}{.5\textwidth}
    \centering
		\includegraphics[width=0.8\textwidth]{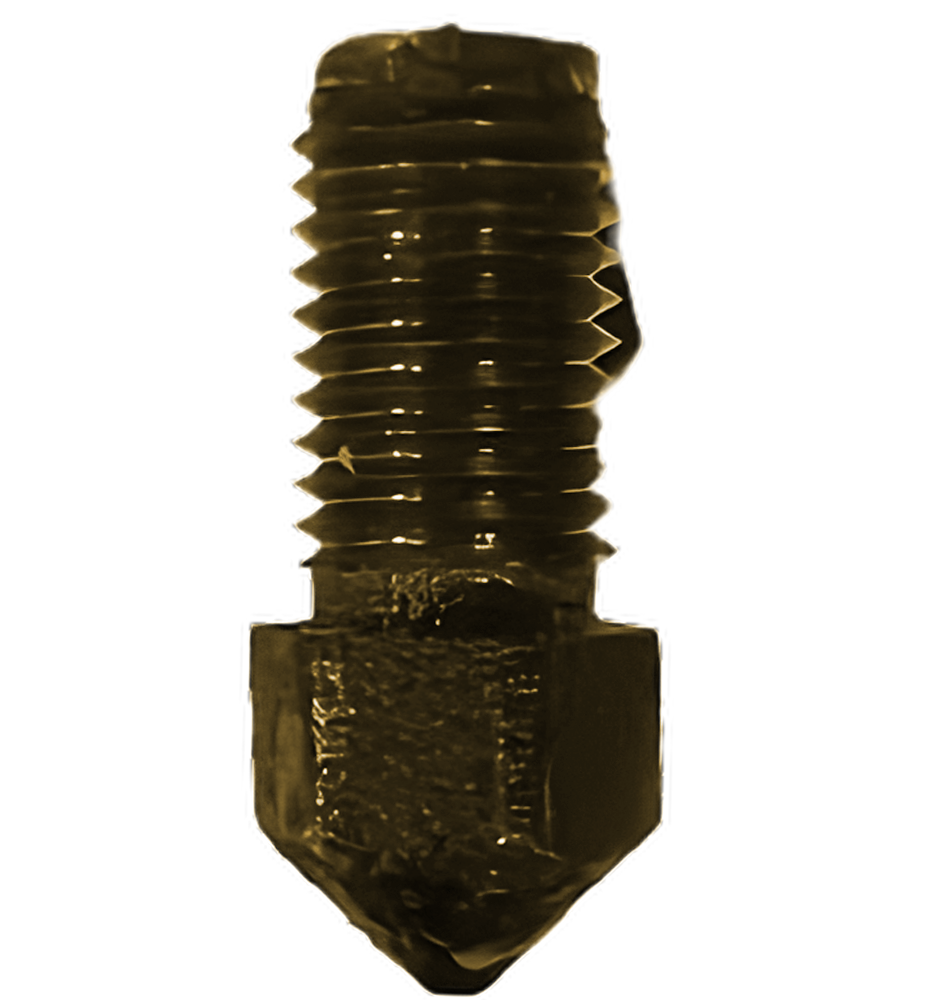}
		\caption{Post-processed nozzle.}
		\label{fig:final_nozzle_1}
\end{subfigure}%
\begin{subfigure}{.5\textwidth}
    \centering
		\includegraphics[width=0.865\textwidth]{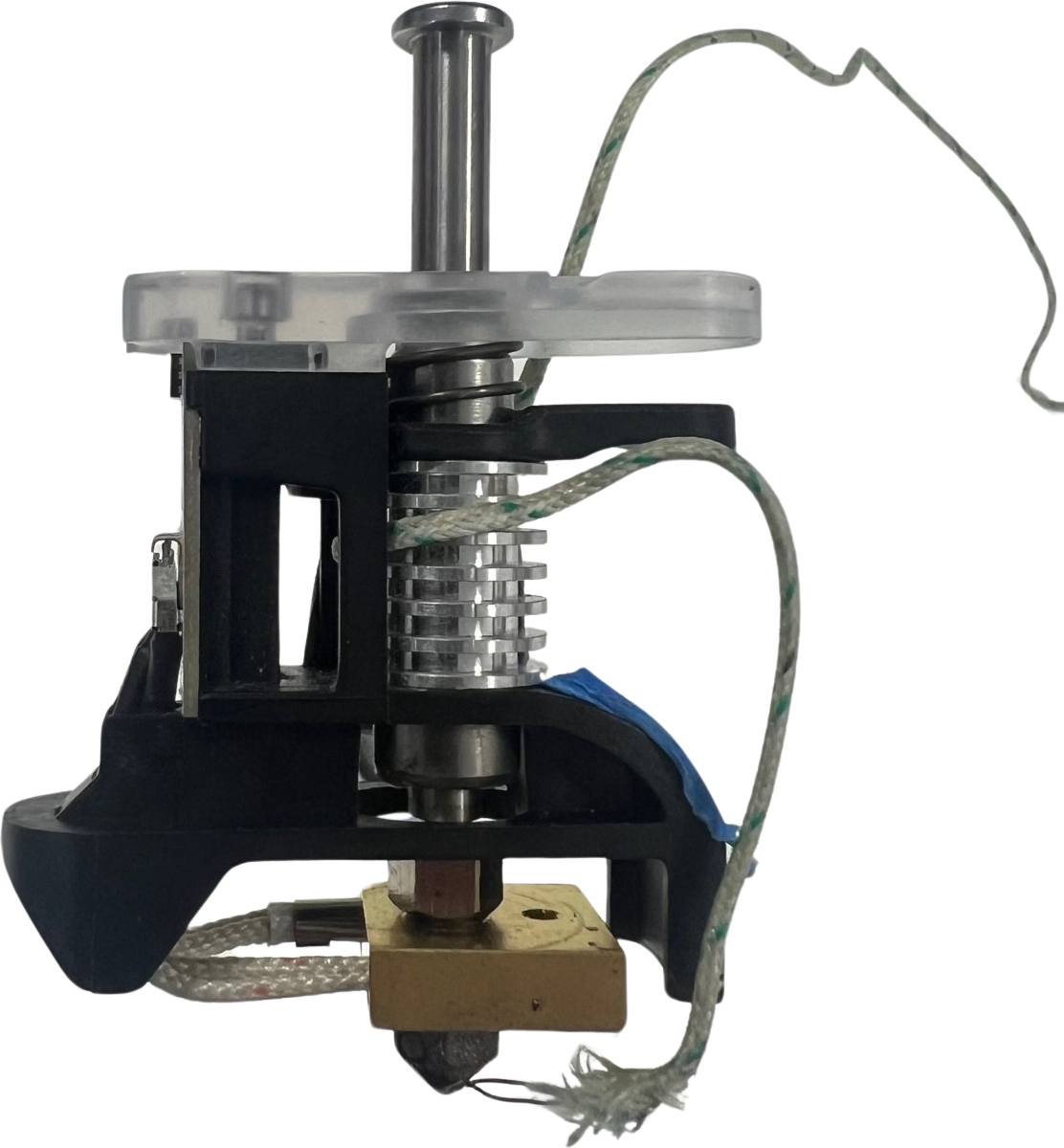}
		\caption{Final nozzle assembled on print-core.}
		\label{fig:final_nozzle_2}
\end{subfigure}
\caption[short]{Final nozzle assembly.}
\label{fig:final_nozzle}
\end{figure}
\begin{figure}[htbp]
	\begin{center}		
		\includegraphics[width=0.6\textwidth]{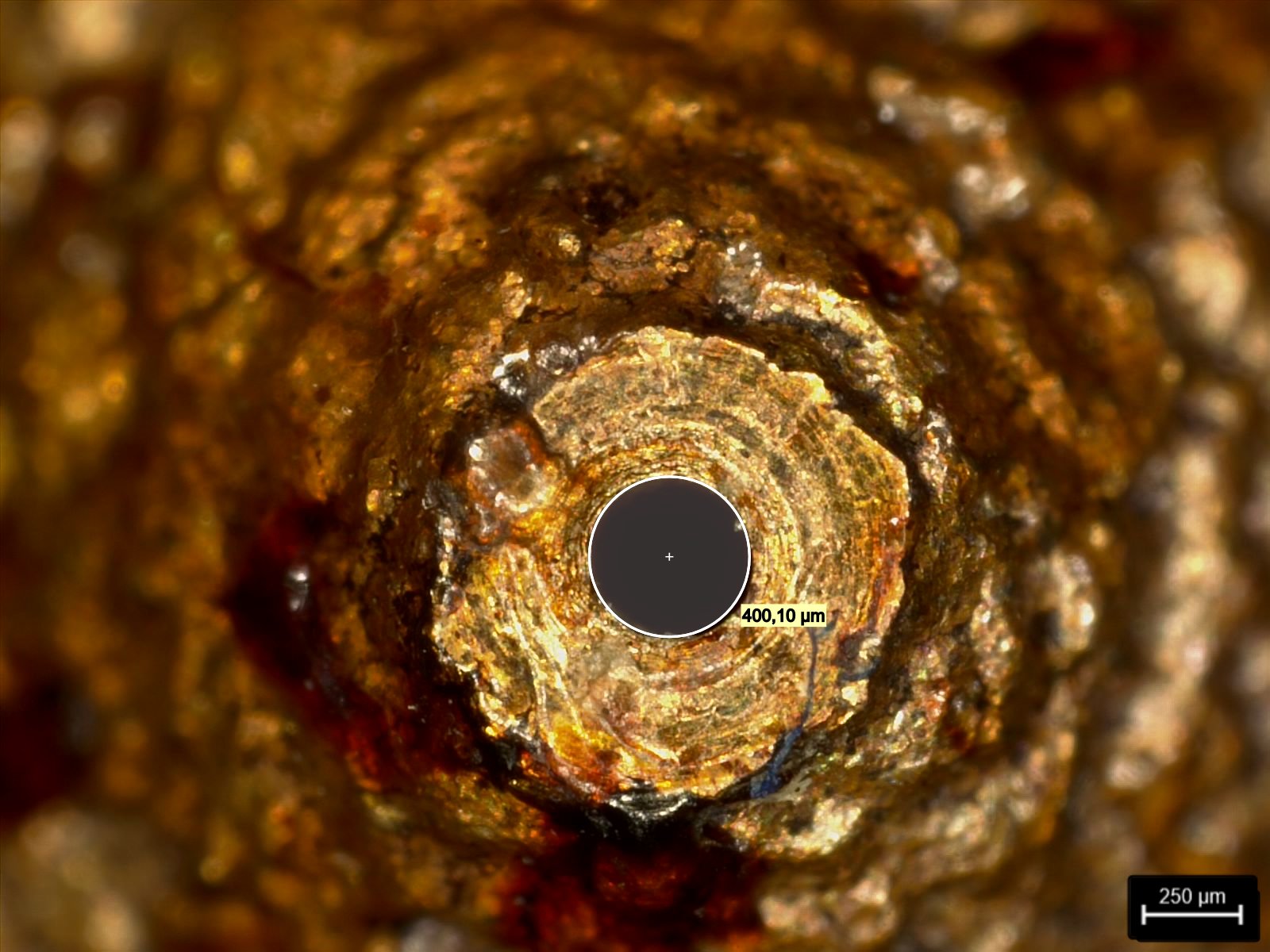}
		\caption{Nozzle die closeup.}
		\label{fig:die_dims}
	\end{center}
\end{figure}

Figure \ref{fig:final_nozzle} presents a finalized nozzle, ready for installation for testing. Figure \ref{fig:die_dims} details the dimensions of the finalized die after the reaming process.

\FloatBarrier
\clearpage
\newpage
 \bibliographystyle{elsarticle-num} 
 \bibliography{myrefs}

\begin{thebibliography}{10}
\expandafter\ifx\csname url\endcsname\relax
  \def\url#1{\texttt{#1}}\fi
\expandafter\ifx\csname urlprefix\endcsname\relax\def\urlprefix{URL }\fi
\expandafter\ifx\csname href\endcsname\relax
  \def\href#1#2{#2} \def\path#1{#1}\fi

\bibitem{OSSWALD201851}
T.~A. Osswald, J.~Puentes, J.~Kattinger, \href{https://www.sciencedirect.com/science/article/pii/S221486041730458X}{{Fused filament fabrication melting model}}, Additive Manufacturing 22 (2018) 51--59.
\newblock \href {https://doi.org/https://doi.org/10.1016/j.addma.2018.04.030} {\path{doi:https://doi.org/10.1016/j.addma.2018.04.030}}.
\newline\urlprefix\url{https://www.sciencedirect.com/science/article/pii/S221486041730458X}

\bibitem{GAO2021101658}
X.~Gao, S.~Qi, X.~Kuang, Y.~Su, J.~Li, D.~Wang, \href{https://www.sciencedirect.com/science/article/pii/S2214860420310307}{{Fused filament fabrication of polymer materials: A review of interlayer bond}}, Additive Manufacturing 37 (2021) 101658.
\newblock \href {https://doi.org/https://doi.org/10.1016/j.addma.2020.101658} {\path{doi:https://doi.org/10.1016/j.addma.2020.101658}}.
\newline\urlprefix\url{https://www.sciencedirect.com/science/article/pii/S2214860420310307}

\bibitem{Yadav2022}
A.~Yadav, P.~Rohru, A.~Babbar, R.~Kumar, N.~Ranjan, J.~S. Chohan, R.~Kumar, M.~Gupta, \href{https://doi.org/10.1007/s12008-022-01026-5}{{Fused filament fabrication: A state-of-the-art review of the technology, materials, properties and defects}}, International Journal on Interactive Design and Manufacturing (IJIDeM) (Aug 2022).
\newblock \href {https://doi.org/10.1007/s12008-022-01026-5} {\path{doi:10.1007/s12008-022-01026-5}}.
\newline\urlprefix\url{https://doi.org/10.1007/s12008-022-01026-5}

\bibitem{turner}
N.~Turner, B.~Strong, S.~Gold, {A review of melt extrusion additive manufacturing processes: I. Process design and modeling}, Rapid Prototyping Journal 20 (2014) 192--204.

\bibitem{turner2}
N.~Turner, S.~Gold, {A review of melt extrusion additive manufacturing processes: II. Materials, dimensional accuracy, and surface roughness}, Rapid Prototyping Journal 21 (2015) 250--261.

\bibitem{larsson_nozzle}
J.~Larsson, P.~Lindstr{\"o}m, C.~Korin, J.~Ekengren, P.~Karlsson, {The Influence of Nozzle Size on the Printing Process and the Mechanical Properties of FFF-Printed Parts}, in: C.~Klahn, M.~Meboldt, J.~Ferchow (Eds.), Industrializing Additive Manufacturing, Springer International Publishing, Cham, 2024, pp. 159--170.

\bibitem{Bikas2016}
H.~Bikas, P.~Stavropoulos, G.~Chryssolouris, \href{https://doi.org/10.1007/s00170-015-7576-2}{{Additive manufacturing methods and modelling approaches: a critical review}}, The International Journal of Advanced Manufacturing Technology 83~(1) (2016) 389--405.
\newblock \href {https://doi.org/10.1007/s00170-015-7576-2} {\path{doi:10.1007/s00170-015-7576-2}}.
\newline\urlprefix\url{https://doi.org/10.1007/s00170-015-7576-2}

\bibitem{POOLE2023105106}
R.~J. Poole, \href{https://www.sciencedirect.com/science/article/pii/S0377025723001180}{{Inelastic and flow-type parameter models for non-Newtonian fluids}}, Journal of Non-Newtonian Fluid Mechanics 320 (2023) 105106.
\newblock \href {https://doi.org/https://doi.org/10.1016/j.jnnfm.2023.105106} {\path{doi:https://doi.org/10.1016/j.jnnfm.2023.105106}}.
\newline\urlprefix\url{https://www.sciencedirect.com/science/article/pii/S0377025723001180}

\bibitem{PEREZCAMACHO2015260}
M.~Pérez-Camacho, J.~López-Aguilar, F.~Calderas, O.~Manero, M.~Webster, \href{https://www.sciencedirect.com/science/article/pii/S0377025715000269}{{Pressure-drop and kinematics of viscoelastic flow through an axisymmetric contraction–expansion geometry with various contraction-ratios}}, Journal of Non-Newtonian Fluid Mechanics 222 (2015) 260--271, rheometry (and General Rheology): Festschrift dedicated to Professor K Walters FRS on the occasion of his 80th birthday.
\newblock \href {https://doi.org/https://doi.org/10.1016/j.jnnfm.2015.01.013} {\path{doi:https://doi.org/10.1016/j.jnnfm.2015.01.013}}.
\newline\urlprefix\url{https://www.sciencedirect.com/science/article/pii/S0377025715000269}

\bibitem{sietsepaper}
S.~de~Vries, T.~Schuller, P.~Fanzio, F.-J. Galindo-Rosales, \href{https://doi.org/10.48550/arXiv.2310.01901}{{Pressure drop non-linearities in Fused Filament Fabrication}} (2023).
\newblock \href {https://doi.org/10.48550/arXiv.2310.01901} {\path{doi:10.48550/arXiv.2310.01901}}.
\newline\urlprefix\url{https://doi.org/10.48550/arXiv.2310.01901}

\bibitem{paper2}
T.~Schuller, P.~Fanzio, F.-J. Galindo-Rosales, \href{https://doi.org/10.48550/arXiv.2311.05158}{{The impact of polymer rheology on the extrusion flow in fused filament fabrication}}, arXiv (2023).
\newblock \href {https://doi.org/10.48550/arXiv.2311.05158} {\path{doi:10.48550/arXiv.2311.05158}}.
\newline\urlprefix\url{https://doi.org/10.48550/arXiv.2311.05158}

\bibitem{rothstein}
J.~Rothstein, G.~McKinley, \href{http://www.sciencedirect.com/science/article/pii/S0377025701000945}{{The axisymmetric contraction–expansion: the role of extensional rheology on vortex growth dynamics and the enhanced pressure drop}}, Journal of Non-Newtonian Fluid Mechanics 98~(1) (2001) 33--63.
\newblock \href {https://doi.org/https://doi.org/10.1016/S0377-0257(01)00094-5} {\path{doi:https://doi.org/10.1016/S0377-0257(01)00094-5}}.
\newline\urlprefix\url{http://www.sciencedirect.com/science/article/pii/S0377025701000945}

\bibitem{rothstein1}
J.~P. Rothstein, G.~H. McKinley, \href{http://www.sciencedirect.com/science/article/pii/S037702579800202X}{{Extensional flow of a polystyrene Boger fluid through a 4:1:4 axisymmetric contraction-expansion}}, Journal of Non-Newtonian Fluid Mechanics 86~(1) (1999) 61--88.
\newblock \href {https://doi.org/https://doi.org/10.1016/S0377-0257(98)00202-X} {\path{doi:https://doi.org/10.1016/S0377-0257(98)00202-X}}.
\newline\urlprefix\url{http://www.sciencedirect.com/science/article/pii/S037702579800202X}

\bibitem{james1982planar}
D.~F. James, J.~H. Saringer, {Planar sink flow of a dilute polymer solution}, Journal of Rheology 26~(3) (1982) 321--325.

\bibitem{nigen2002viscoelastic}
S.~Nigen, K.~Walters, {Viscoelastic contraction flows: comparison of axisymmetric and planar configurations}, Journal of non-newtonian fluid mechanics 102~(2) (2002) 343--359.

\bibitem{rodd2005inertio}
L.~E. Rodd, T.~P. Scott, D.~V. Boger, J.~J. Cooper-White, G.~H. McKinley, {The inertio-elastic planar entry flow of low-viscosity elastic fluids in micro-fabricated geometries}, Journal of Non-Newtonian Fluid Mechanics 129~(1) (2005) 1--22.

\bibitem{campo2011flow}
L.~Campo-Dea{\~n}o, F.~J. Galindo-Rosales, F.~T. Pinho, M.~A. Alves, M.~S. Oliveira, {Flow of low viscosity Boger fluids through a microfluidic hyperbolic contraction}, Journal of Non-Newtonian Fluid Mechanics 166~(21-22) (2011) 1286--1296.

\bibitem{schuller_add_ma}
T.~Schuller, P.~Fanzio, F.-J. Galindo-Rosales, \href{https://www.sciencedirect.com/science/article/pii/S2214860422003463}{{Analysis of the importance of shear-induced elastic stresses in material extrusion}}, Additive Manufacturing 57 (2022).
\newblock \href {https://doi.org/https://doi.org/10.1016/j.addma.2022.102952} {\path{doi:https://doi.org/10.1016/j.addma.2022.102952}}.
\newline\urlprefix\url{https://www.sciencedirect.com/science/article/pii/S2214860422003463}

\bibitem{james2021}
D.~F. James, C.~A. Roos, \href{https://www.sciencedirect.com/science/article/pii/S0377025721000707}{{Pressure drop of a Boger fluid in a converging channel}}, Journal of Non-Newtonian Fluid Mechanics 293 (2021) 104557.
\newblock \href {https://doi.org/https://doi.org/10.1016/j.jnnfm.2021.104557} {\path{doi:https://doi.org/10.1016/j.jnnfm.2021.104557}}.
\newline\urlprefix\url{https://www.sciencedirect.com/science/article/pii/S0377025721000707}

\bibitem{rao2019engineering}
S.~Rao, \href{https://books.google.pt/books?id=oG21DwAAQBAJ}{{Engineering Optimization: Theory and Practice}}, Wiley, 2019.
\newline\urlprefix\url{https://books.google.pt/books?id=oG21DwAAQBAJ}

\bibitem{siddall1982optimal}
J.~Siddall, \href{https://books.google.pt/books?id=i2hyniQpecYC}{{Optimal Engineering Design: Principles and Applications}}, Dekker Mechanical Engineering, Taylor \& Francis, 1982.
\newline\urlprefix\url{https://books.google.pt/books?id=i2hyniQpecYC}

\bibitem{mdo_book}
J.~Martins, A.~Ning, {Engineering Design Optimization}, 2021.
\newblock \href {https://doi.org/10.1017/9781108980647} {\path{doi:10.1017/9781108980647}}.

\bibitem{opt_cca}
C.~C. António, {Otimização de Sistemas em Engenharia - Fundamentos e Algoritmos para o Projeto Óptimo}, Engebook, 2020.

\bibitem{Galindo-Rosales2023}
F.~J. Galindo-Rosales, {Shear Thickening Fluid/Cork Composites Against Blunt Impacts in Football Shin Guards Applications}, Springer International Publishing, Cham, 2023, pp. 41--61.
\newblock \href {https://doi.org/10.1007/978-3-031-35521-9\_4} {\path{doi:10.1007/978-3-031-35521-9\_4}}.

\bibitem{ma12071086}
T.~Rodrigues, F.~J. Galindo-Rosales, L.~Campo-Deaño, \href{https://www.mdpi.com/1996-1944/12/7/1086}{{Towards an Optimal Pressure Tap Design for Fluid-Flow Characterisation at Microscales}}, Materials 12~(7) (2019).
\newblock \href {https://doi.org/10.3390/ma12071086} {\path{doi:10.3390/ma12071086}}.
\newline\urlprefix\url{https://www.mdpi.com/1996-1944/12/7/1086}

\bibitem{C3RA47230B}
F.-J. Galindo-Rosales, M.~S.~N. Oliveira, M.~A. Alves, \href{http://dx.doi.org/10.1039/C3RA47230B}{{Optimized cross-slot microdevices for homogeneous extension}}, RSC Adv. 4 (2014) 7799--7804.
\newblock \href {https://doi.org/10.1039/c3ra47230b} {\path{doi:10.1039/c3ra47230b}}.
\newline\urlprefix\url{http://dx.doi.org/10.1039/C3RA47230B}

\bibitem{borrvall_petersson}
T.~Borrvall, J.~Petersson, {Topology optimization of fluids in Stokes flow}, International Journal for Numerical Methods in Fluids 41~(1) (2003) 77--107.
\newblock \href {https://doi.org/10.1002/fld.426} {\path{doi:10.1002/fld.426}}.

\bibitem{JensenAPL2012}
K.~Ejlebjerg~Jensen, P.~Szabo, F.~Okkels, \href{https://doi.org/10.1063/1.4728108}{{Topology optimization of viscoelastic rectifiers}}, Applied Physics Letters 100~(23) (2012) 234102.
\newblock \href {http://arxiv.org/abs/https://doi.org/10.1063/1.4728108} {\path{arXiv:https://doi.org/10.1063/1.4728108}}, \href {https://doi.org/10.1063/1.4728108} {\path{doi:10.1063/1.4728108}}.
\newline\urlprefix\url{https://doi.org/10.1063/1.4728108}

\bibitem{Groisman2004}
A.~Groisman, S.~R. Quake, \href{https://doi.org/10.1103/physrevlett.92.094501}{{A Microfluidic Rectifier: Anisotropic Flow Resistance at Low Reynolds Numbers}}, Physical Review Letters 92~(9) (Mar. 2004).
\newblock \href {https://doi.org/10.1103/physrevlett.92.094501} {\path{doi:10.1103/physrevlett.92.094501}}.
\newline\urlprefix\url{https://doi.org/10.1103/physrevlett.92.094501}

\bibitem{SOUSA2010652}
P.~Sousa, F.~Pinho, M.~Oliveira, M.~Alves, \href{https://www.sciencedirect.com/science/article/pii/S0377025710000911}{{Efficient microfluidic rectifiers for viscoelastic fluid flow}}, Journal of Non-Newtonian Fluid Mechanics 165~(11) (2010) 652--671.
\newblock \href {https://doi.org/https://doi.org/10.1016/j.jnnfm.2010.03.005} {\path{doi:https://doi.org/10.1016/j.jnnfm.2010.03.005}}.
\newline\urlprefix\url{https://www.sciencedirect.com/science/article/pii/S0377025710000911}

\bibitem{pimenta4}
F.~Pimenta, R.~G. Sousa, M.~A. Alves, \href{https://doi.org/10.1063/1.5037472}{{Optimization of flow-focusing devices for homogeneous extensional flow}}, Biomicrofluidics 12~(5) (2018) 054103.
\newblock \href {http://arxiv.org/abs/https://doi.org/10.1063/1.5037472} {\path{arXiv:https://doi.org/10.1063/1.5037472}}, \href {https://doi.org/10.1063/1.5037472} {\path{doi:10.1063/1.5037472}}.
\newline\urlprefix\url{https://doi.org/10.1063/1.5037472}

\bibitem{Zografosbiomic2016}
K.~Zografos, F.~Pimenta, M.~A. Alves, M.~S.~N. Oliveira, \href{https://doi.org/10.1063/1.4954814}{{Microfluidic converging/diverging channels optimised for homogeneous extensional deformation}}, Biomicrofluidics 10~(4) (2016) 043508.
\newblock \href {http://arxiv.org/abs/https://doi.org/10.1063/1.4954814} {\path{arXiv:https://doi.org/10.1063/1.4954814}}, \href {https://doi.org/10.1063/1.4954814} {\path{doi:10.1063/1.4954814}}.
\newline\urlprefix\url{https://doi.org/10.1063/1.4954814}

\bibitem{zografos2019}
K.~Zografos, S.~J. Haward, M.~S.~N. Oliveira, \href{https://doi.org/10.1007/s10404-019-2295-x}{{Optimised multi-stream microfluidic designs for controlled extensional deformation}}, Microfluidics and Nanofluidics 23~(12) (Nov. 2019).
\newblock \href {https://doi.org/10.1007/s10404-019-2295-x} {\path{doi:10.1007/s10404-019-2295-x}}.
\newline\urlprefix\url{https://doi.org/10.1007/s10404-019-2295-x}

\bibitem{Haward2012}
S.~J. Haward, M.~S.~N. Oliveira, M.~A. Alves, G.~H. McKinley, \href{https://doi.org/10.1103/physrevlett.109.128301}{{Optimized Cross-Slot Flow Geometry for Microfluidic Extensional Rheometry}}, Physical Review Letters 109~(12) (Sep. 2012).
\newblock \href {https://doi.org/10.1103/physrevlett.109.128301} {\path{doi:10.1103/physrevlett.109.128301}}.
\newline\urlprefix\url{https://doi.org/10.1103/physrevlett.109.128301}

\bibitem{liu_zografos}
Y.~Liu, K.~Zografos, J.~Fidalgo, C.~Duch\^ene, C.~Quintard, T.~Darnige, V.~Filipe, S.~Huille, O.~d. Roure, M.~S.~N. Oliveira, A.~Lindner, \href{https://arxiv.org/abs/2008.01134}{{Optimised hyperbolic microchannels for the mechanical characterisation of bio-particles}} (2020).
\newblock \href {https://doi.org/10.48550/arxiv.2008.01134} {\path{doi:10.48550/arxiv.2008.01134}}.
\newline\urlprefix\url{https://arxiv.org/abs/2008.01134}

\bibitem{adams2020dakota}
B.~M. Adams~et al., {Dakota, A Multilevel Parallel Object-Oriented Framework for Design Optimization, Parameter Estimation, Uncertainty Quantification, and Sensitivity Analysis: Version 6.12 Reference Manual}, 2020.

\bibitem{foam_jasak}
H.~G. Weller, G.~Tabor, H.~Jasak, C.~Fureby, \href{https://aip.scitation.org/doi/abs/10.1063/1.168744}{{A tensorial approach to computational continuum mechanics using object-oriented techniques}}, Computers in Physics 12~(6) (1998) 620--631.
\newblock \href {http://arxiv.org/abs/https://aip.scitation.org/doi/pdf/10.1063/1.168744} {\path{arXiv:https://aip.scitation.org/doi/pdf/10.1063/1.168744}}, \href {https://doi.org/10.1063/1.168744} {\path{doi:10.1063/1.168744}}.
\newline\urlprefix\url{https://aip.scitation.org/doi/abs/10.1063/1.168744}

\bibitem{Geuzaine2009}
C.~Geuzaine, J.-F. Remacle, \href{https://doi.org/10.1002/nme.2579}{{Gmsh: A 3-D finite element mesh generator with built-in pre- and post-processing facilities}}, International Journal for Numerical Methods in Engineering 79~(11) (2009) 1309--1331.
\newblock \href {https://doi.org/10.1002/nme.2579} {\path{doi:10.1002/nme.2579}}.
\newline\urlprefix\url{https://doi.org/10.1002/nme.2579}

\bibitem{Li2020}
M.~Li, W.~Du, A.~Elwany, Z.~Pei, C.~Ma, \href{https://doi.org/10.1115/1.4047430}{{Metal Binder Jetting Additive Manufacturing: A Literature Review}}, Journal of Manufacturing Science and Engineering 142~(9) (Jun. 2020).
\newblock \href {https://doi.org/10.1115/1.4047430} {\path{doi:10.1115/1.4047430}}.
\newline\urlprefix\url{https://doi.org/10.1115/1.4047430}

\bibitem{Huang1991}
D.~Huang, E.~A. Swanson, C.~P. Lin, J.~S. Schuman, W.~G. Stinson, W.~Chang, M.~R. Hee, T.~Flotte, K.~Gregory, C.~A. Puliafito, J.~G. Fujimoto, \href{https://doi.org/10.1126/science.1957169}{{Optical Coherence Tomography}}, Science 254~(5035) (1991) 1178--1181.
\newblock \href {https://doi.org/10.1126/science.1957169} {\path{doi:10.1126/science.1957169}}.
\newline\urlprefix\url{https://doi.org/10.1126/science.1957169}

\bibitem{vanderKolk2023}
J.~van~der Kolk, D.~Tieman, M.~Jalaal, \href{https://doi.org/10.1017/jfm.2023.118}{{Viscoplastic lines: printing a single filament of yield stress material on a surface}}, Journal of Fluid Mechanics 958 (Mar. 2023).
\newblock \href {https://doi.org/10.1017/jfm.2023.118} {\path{doi:10.1017/jfm.2023.118}}.
\newline\urlprefix\url{https://doi.org/10.1017/jfm.2023.118}

\bibitem{ORTEGACASANOVA2019102}
J.~Ortega-Casanova, M.~Jimenez-Canet, F.~Galindo-Rosales, \href{https://www.sciencedirect.com/science/article/pii/S0017931019310737}{{Numerical study of the heat and momentum transfer between a flat plate and an impinging jet of power law fluids}}, International Journal of Heat and Mass Transfer 141 (2019) 102--111.
\newblock \href {https://doi.org/https://doi.org/10.1016/j.ijheatmasstransfer.2019.06.072} {\path{doi:https://doi.org/10.1016/j.ijheatmasstransfer.2019.06.072}}.
\newline\urlprefix\url{https://www.sciencedirect.com/science/article/pii/S0017931019310737}

\bibitem{james2016}
D.~F. James, \href{https://www.sciencedirect.com/science/article/pii/S037702571600015X}{{N1 stresses in extensional flows}}, Journal of Non-Newtonian Fluid Mechanics 232 (2016) 33--42.
\newblock \href {https://doi.org/https://doi.org/10.1016/j.jnnfm.2016.01.012} {\path{doi:https://doi.org/10.1016/j.jnnfm.2016.01.012}}.
\newline\urlprefix\url{https://www.sciencedirect.com/science/article/pii/S037702571600015X}

\bibitem{Walters2009TheCR}
K.~Walters, H.~R. Tamaddon-Jahromi, M.~F. Webster, M.~F. Tom{\'e}, S.~McKee, \href{https://api.semanticscholar.org/CorpusID:118748112}{The competing roles of extensional viscosity and normal stress differences in complex flows of elastic liquids}, Korea-australia Rheology Journal 21 (2009) 225--233.
\newline\urlprefix\url{https://api.semanticscholar.org/CorpusID:118748112}

\bibitem{Tseng2023}
H.-C. Tseng, \href{https://doi.org/10.1007/s13367-023-00070-1}{The competing role of shear and extension-induced first normal stress differences within a mixed flow for a viscoelastic fluid}, Korea-Australia Rheology Journal (Sep 2023).
\newblock \href {https://doi.org/10.1007/s13367-023-00070-1} {\path{doi:10.1007/s13367-023-00070-1}}.
\newline\urlprefix\url{https://doi.org/10.1007/s13367-023-00070-1}

\bibitem{giesekus}
H.~Giesekus, \href{http://www.sciencedirect.com/science/article/pii/0377025782850167}{{A simple constitutive equation for polymer fluids based on the concept of deformation-dependent tensorial mobility}}, Journal of Non-Newtonian Fluid Mechanics 11~(1) (1982) 69--109.
\newblock \href {https://doi.org/https://doi.org/10.1016/0377-0257(82)85016-7} {\path{doi:https://doi.org/10.1016/0377-0257(82)85016-7}}.
\newline\urlprefix\url{http://www.sciencedirect.com/science/article/pii/0377025782850167}

\bibitem{GIESEKUS1985349}
H.~Giesekus, \href{https://www.sciencedirect.com/science/article/pii/0377025785800264}{{Constitutive equations for polymer fluids based on the concept of configuration-dependent molecular mobility: a generalized mean-configuration model}}, Journal of Non-Newtonian Fluid Mechanics 17~(3) (1985) 349--372.
\newblock \href {https://doi.org/https://doi.org/10.1016/0377-0257(85)80026-4} {\path{doi:https://doi.org/10.1016/0377-0257(85)80026-4}}.
\newline\urlprefix\url{https://www.sciencedirect.com/science/article/pii/0377025785800264}

\bibitem{rheoTool}
F.~Pimenta, M.~Alves, {rheoTool}, \url{https://github.com/fppimenta/rheoTool} (2016).

\bibitem{pimenta1}
F.~Pimenta, M.~Alves, \href{http://www.sciencedirect.com/science/article/pii/S0377025716303329}{{Stabilization of an open-source finite-volume solver for viscoelastic fluid flows}}, Journal of Non-Newtonian Fluid Mechanics 239 (2017) 85--104.
\newblock \href {https://doi.org/https://doi.org/10.1016/j.jnnfm.2016.12.002} {\path{doi:https://doi.org/10.1016/j.jnnfm.2016.12.002}}.
\newline\urlprefix\url{http://www.sciencedirect.com/science/article/pii/S0377025716303329}

\bibitem{AlvesetalAnnRevFluMech2021}
M.~Alves, P.~Oliveira, F.~Pinho, \href{https://doi.org/10.1146/annurev-fluid-010719-060107}{{Numerical Methods for Viscoelastic Fluid Flows}}, Annual Review of Fluid Mechanics 53~(1) (2021) 509--541.
\newblock \href {http://arxiv.org/abs/https://doi.org/10.1146/annurev-fluid-010719-060107} {\path{arXiv:https://doi.org/10.1146/annurev-fluid-010719-060107}}, \href {https://doi.org/10.1146/annurev-fluid-010719-060107} {\path{doi:10.1146/annurev-fluid-010719-060107}}.
\newline\urlprefix\url{https://doi.org/10.1146/annurev-fluid-010719-060107}

\bibitem{nasaGCS}
Nasa, {Examining Spatial (Grid) Convergence}, https://www.grc.nasa.gov/www/wind/valid/tutorial/spatconv.html, last Updated: Wednesday, 10-Feb-2021 09:38:59 EST.

\bibitem{Roache1972}
P.~J. Roache, { Computational fluid dynamics}, Hermosa Publishers Albuquerque, N.M, 1972.

\bibitem{Roache2006-dc}
P.~J. Roache, {Verification and Validation in Computational Science and Engineering}, Hermosa, 2006.

\bibitem{Roache1993}
P.~J. Roache, P.~M. Knupp, \href{https://doi.org/10.1002/cnm.1640090502}{{Completed Richardson extrapolation}}, Communications in Numerical Methods in Engineering 9~(5) (1993) 365--374.
\newblock \href {https://doi.org/10.1002/cnm.1640090502} {\path{doi:10.1002/cnm.1640090502}}.
\newline\urlprefix\url{https://doi.org/10.1002/cnm.1640090502}

\bibitem{1911}
L.~F. Richardson, \href{https://doi.org/10.1098/rsta.1911.0009}{{IX. The approximate arithmetical solution by finite differences of physical problems involving differential equations, with an application to the stresses in a masonry dam}}, Philosophical Transactions of the Royal Society of London. Series A, Containing Papers of a Mathematical or Physical Character 210~(459-470) (1911) 307--357.
\newblock \href {https://doi.org/10.1098/rsta.1911.0009} {\path{doi:10.1098/rsta.1911.0009}}.
\newline\urlprefix\url{https://doi.org/10.1098/rsta.1911.0009}

\bibitem{1927}
L.~F. Richardson, J.~A. Gaunt, \href{https://doi.org/10.1098/rsta.1927.0008}{{VIII. The deferred approach to the limit}}, Philosophical Transactions of the Royal Society of London. Series A, Containing Papers of a Mathematical or Physical Character 226~(636-646) (1927) 299--361.
\newblock \href {https://doi.org/10.1098/rsta.1927.0008} {\path{doi:10.1098/rsta.1927.0008}}.
\newline\urlprefix\url{https://doi.org/10.1098/rsta.1927.0008}

\bibitem{foamscripts}
T.~Schuller, {foamScripts}, \url{https://github.com/T-Schuller/foamScripts} (2023).

\end{thebibliography}





\end{document}